\documentclass[aps,reprint,prb,amsfonts,amsmath,superscriptaddress]{revtex4-2}

\usepackage{graphicx}
\usepackage{times}
\usepackage{hyperref}
\usepackage[caption=false]{subfig}

\begin{document}

\title{Weak first-order phase transitions in the frustrated square lattice $J_1$-$J_2$ classical Ising model}

\author{Adil A. Gangat}
\affiliation{Department of Physics, National Taiwan University, Taipei 10607, Taiwan}
\affiliation{Département de Physique and Institut Quantique, Université de Sherbrooke, Sherbrooke, Québec, Canada J1K 2R1}
\affiliation{School of Physics, Georgia Institute of Technology, Atlanta, GA 30332}
\affiliation{ARC Centre of Excellence for Engineered Quantum Systems, School of Mathematics and Physics,  The University of Queensland, St. Lucia, Queensland 4072, Australia}
\affiliation{Physics \& Informatics Laboratories, NTT Research, Inc., Sunnyvale, CA 94085}
\affiliation{Division of Chemistry and Chemical Engineering, California Institute of Technology, Pasadena, CA 91125}
\date{\today}
\begin{abstract}
The classical $J_1$-$J_2$ Ising model on the square lattice is a minimal model of frustrated magnetism whose phase boundaries have remained under scrutiny for decades.  Signs of first-order phase transitions have appeared in some studies, but strong evidence remains lacking.  The current consensus, based upon the numerical data and theoretical arguments in [S. Jin et al., Phys. Rev. Lett. \textbf{108}, 045702 (2012)], is that first-order phase transitions are ruled out in the region  $g = J_2/|J_1|\gtrsim 0.67$. We point out a loophole in the basis for this consensus, and we find strong evidence that the phase boundary is instead weak first-order at $0.67\lesssim g<\infty$ such that it asymptotically becomes second-order when $g\rightarrow\infty$.  We also find strong evidence that the phase boundary is first-order in the region $0.5<g\lesssim0.67$.  We establish these results with adiabatic evolution of matrix product states directly in the thermodynamic limit, and with the theory of finite entanglement scaling.  We also find suggestive evidence that when $g\rightarrow0.5^+$, the phase boundary becomes of an anomalous first-order type wherein the correlation length is very large in one of the coexisting phases but very small in the other. \end{abstract}
\maketitle

\section{Introduction}
\label{sec:intro}
The study of minimal models is well known to provide valuable fundamental insights. A seminal example is the understanding of thermal phase transitions gleaned from the classical Ising model on the square lattice with only nearest-neighbor interactions.  Adding next-nearest-neighbor (i.e., diagonal) interactions to that model provides us with a minimal model of frustrated magnetism. Its Hamiltonian is given by
\begin{equation}
H = J_1\sum_{\langle i,j \rangle} \sigma_i\sigma_j+ J_2\sum_{\langle\langle k,l \rangle\rangle}\sigma_k \sigma_l,
\label{eqn:Ham}
\end{equation}
where $\langle . \rangle$ and $\langle\langle . \rangle\rangle$ respectively denote nearest and next-nearest-neighbors, and $\sigma_i = \pm1$.
Frustration arises when $J_2>0$: the nearest-neighbor and next-nearest-neighbor terms can not be simultaneously minimized on any minimal plaquette of the lattice (regardless of the sign of $J_1$).  While the phases of this model are well understood, a complete understanding of the order of its phase boundaries has eluded decades of effort \cite{nauenberg1974critical,van1975singularities,nightingale1977non,barber1979non,oitmaa1981square,swendsen1979monte,landau1980phase,binder1980phase,landau1985phase,aguilera1993specific,moran1993first,moran1994phase,buzano1997cluster,lopez1999cluster,malakis2006monte,monroe2007phase,dos2008phase,kalz2008phase,kalz2009monte,kalz2011analysis,murtazaev2011critical,murtazaev2013phase,jin2012ashkin,jin2013phase,kalz2012location,murtazaev2015critical,bobak2015phase,ramazanov2016thermodynamic,li2021tensor,hu2021numerical,abalmasov2023metastable,yoshiyama2023higher,abalmasov2023free} (see Refs. \onlinecite{li2021tensor,hu2021numerical} for recent summaries).

\begin{figure}[b]
\includegraphics[scale=0.44]{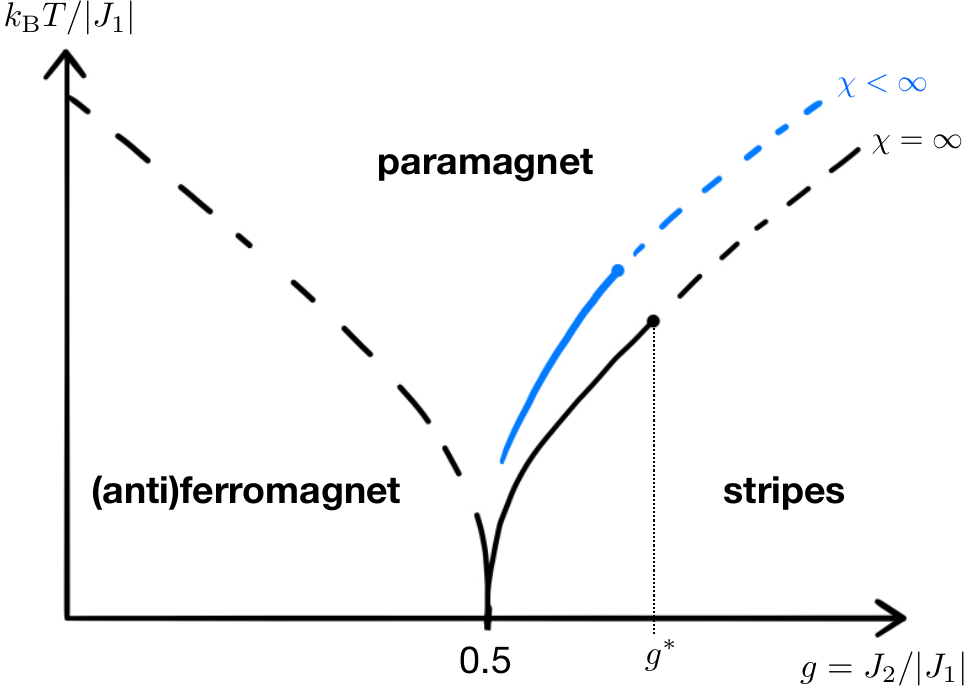}
\caption{(color online). Phase diagram for the model in Eq. (\ref{eqn:Ham}) when $J_2>0$.  The ordered phase when $g>0.5$ has stripes of single-spin width.  The phase transition at $g=0.5$ occurs at zero temperature \cite{hu2021numerical,lee2024frustrated}.  We find that in the region $g>0.5$ there is a value of $g$, denoted $g^*$, for which the phase boundary switches between first-order (solid) and second-order (dashed). We also find that the phase boundary at $g>0.5$ shifts to larger values of $g$ and smaller values of $T$ as $\chi$ (a refinement parameter of matrix product states (MPSs)) is increased. From theoretical arguments, this implies that detecting a first-order phase transition in the region $g>0.5$ with a finite-$\chi$ MPS provides a lower bound on the true value of $g^*$.  This leads us to conclude that $g^*=\infty$, in contradiction to the current consensus that $0.5\leq g^*\lesssim0.67$.
}
\label{fig:phasediagram}
\end{figure}

The phase diagram is illustrated in Fig. \ref{fig:phasediagram}).  The phase transition at $g=0.5$ occurs at zero temperature \cite{hu2021numerical,lee2024frustrated}.  The low temperature phase is either ferromagnetic ($J_1<0$) or antiferromagnetic ($J_1>0$) when $0\leq g < 0.5$, and stripe-ordered when $0.5 < g < \infty$, where $g=J_2/|J_1|$.  Both low temperature phases transition directly to a paramagnetic phase.  Thus, this model provides an opportunity to study a phase transition between striped and paramagnetic phases.

Regarding $g>0.5$, the phase boundary must go to Ising universality when $g\rightarrow\infty$, where the model becomes two decoupled copies of the nearest-neighbor square lattice Ising model.  The current consensus, based on Refs. [\onlinecite{jin2012ashkin,jin2013phase,kalz2012location,murtazaev2015critical}], is that of second-order transitions at $g\gtrsim0.67$.  The order of the transitions at $0.5\leq g\lesssim 0.67$ remains uncertain; there is suggestive evidence that at least part of it is weak first-order \cite{moran1993first,moran1994phase,lopez1999cluster,dos2008phase,kalz2008phase,kalz2009monte,kalz2011analysis,jin2013phase,bobak2015phase,hu2021numerical,yoshiyama2023higher}, but also suggestive evidence that it is entirely second order \cite{malakis2006monte,murtazaev2013phase}.  

Regarding $g<0.5$, the phase boundary must go to Ising universality when $g\rightarrow0^+$.  A few studies find suggestive evidence for first-order transitions in a contiguous range of $g$ up to $0.5$ \cite{hu2021numerical,jin2013phase,bobak2015phase,abalmasov2023metastable,abalmasov2023free}, but strong results are lacking.

\begin{figure*}
\includegraphics[width=\textwidth]{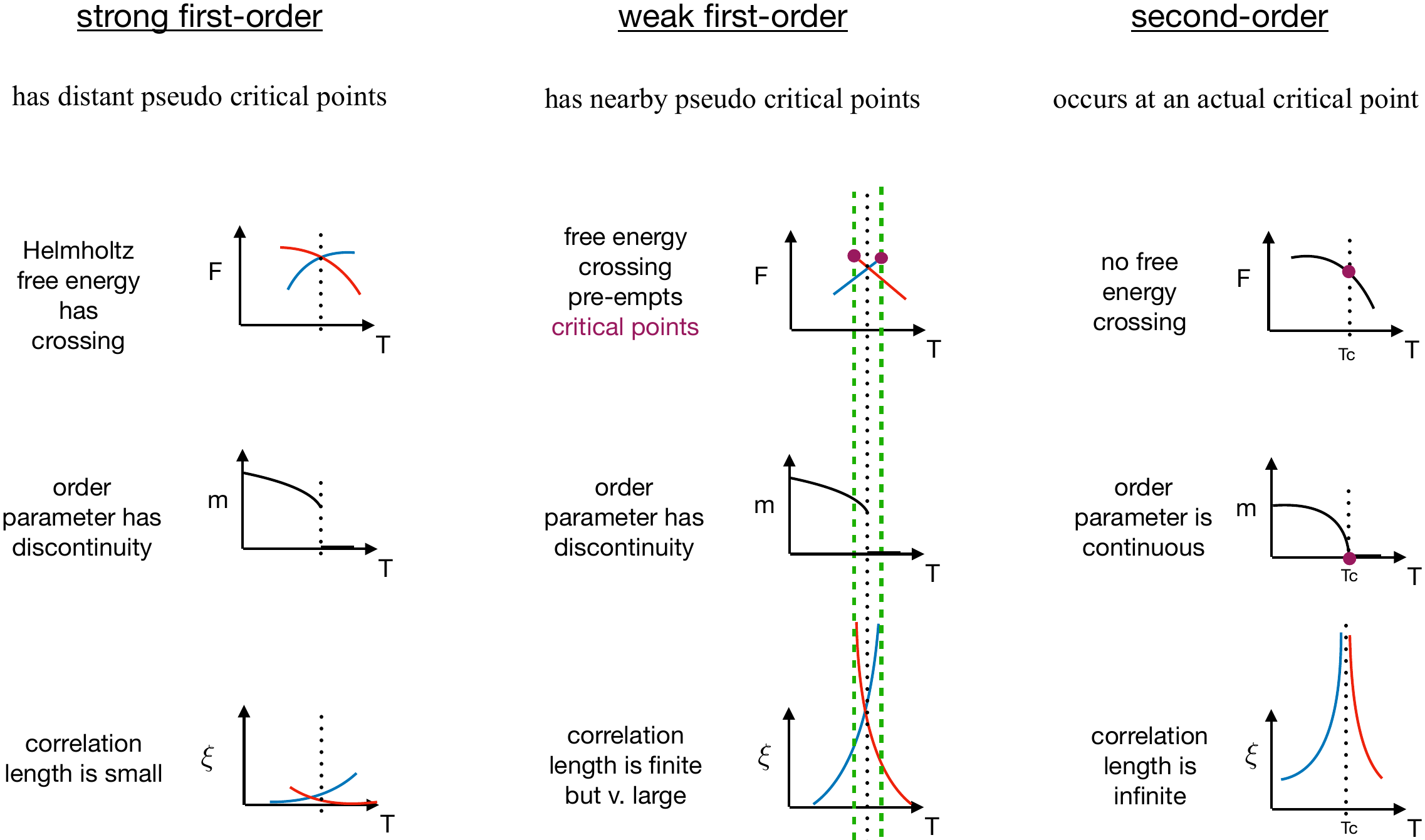}
\caption{(color online). Comparison of strong first-order, weak first-order, and second-order thermal phase transitions. Black dotted lines denote the location of each phase transition, red circles represent critical points, $\xi$ is the correlation length in the equilibrium or metastable phase, $m$ is the order parameter, $T$ is the temperature, and $F=-k_B T~\textrm{ln}~Z$ is the Helmholtz free energy where $k_B$ is Boltzmann's constant and $Z$ is the partition function.  Both strong and weak first-order phase transitions entail a discontinuity in the first derivative of $F$ and a discontinuity in $m$.  However, in a weak first-order phase transition the correlation length in either phase is very large due to the proximity of the pseudo critical points associated with the metastable branches.  It is not necessarily the case, however, that the discontinuity in $m$ is very small at a weak first-order phase transition \cite{igloi1999boundary}.
}
\label{fig:phasetransitions}
\end{figure*}

The difficulty in determining the order of the phase transitions in this model is essentially due to the difficulty in numerically distinguishing between weak first-order and second order phase transitions.  We illuminate the salient features of weak first-order phase transitions that give rise to this difficulty via a comparison between strong and weak first-order phase transitions: Both strong and weak first-order phase transitions necessarily entail a discontinuity in the first derivative of the Helmholtz free energy and a discontinuity of the order parameter (see Fig. \ref{fig:phasetransitions}).  Both also entail metastable branches of finite extent (terminating at \textit{spinodal points}) in the phase diagram.  As a spinodal point is approached along a metastable branch, the metastable state approaches criticality as if there were a critical point (termed a \textit{pseudo critical point} \cite{fernandez1992weak,schulke2000dynamic} \footnote{This usage of ``pseudo critical point'' is different from the usage in Ref. \cite{tagliacozzo2008scaling}.} or \textit{pseudospinodal point} \cite{binder1987theory}) located beyond (or, in the mean-field limit, precisely at) the spinodal point \cite{binder1987theory}.  Around a weak first-order phase transition, these pseudo critical points lie very close to the phase transition, which makes the coexistence region very small and imparts a large (but still finite) correlation length to the phase transition.  Around a strong first-order phase transition, the pseudo critical points are far from the phase transition, and the correlation lengths are therefore small.  For a weak first-order phase transition, the smallness of the coexistence region means that the hysteresis is difficult to resolve, and the large correlation lengths can make weak first-order phase transitions numerically appear as second-order phase transitions \cite{iino2019detecting}.

For the region $0<g\lesssim0.67$ in the present model, all existing results that show first-order signatures leave open the possibility that those first-order signatures are only numerical artifacts (due to, for example, finite cluster size), and all existing results that show second-order signatures leave open the possibility that those second-order signatures are in reality due to weak first-order phase transitions.  For the region $g\gtrsim0.67$, the same is true for all results that appear prior to Ref. \cite{jin2012ashkin}.  The authors of Ref. \cite{jin2012ashkin} (seemingly) make a breakthrough by combining numerical data with theoretical arguments instead of relying upon numerical data alone.  Specifically, they (seemingly) rule out the possibility of first-order transitions at $g\gtrsim0.67$ in the following way: because the present model has a theoretically known second-order phase transition at $g=\infty$, they \cite{jin2012ashkin} assume that there is a line of second-order transitions that connects to it, and they combine numerical data with theoretical arguments to conclude that this line of second-order transitions in the region $g\gtrsim0.67$ maps on to a phase boundary of the Ashkin-Teller model that exhibits continuously varying critical exponents. Subsequent numerics \cite{jin2013phase,kalz2012location,murtazaev2015critical} confirm signatures of this ``Ashkin-Teller criticality" at $g\gtrsim0.67$, hence the current consensus.

\begin{figure}[t]
\includegraphics[scale=0.8]{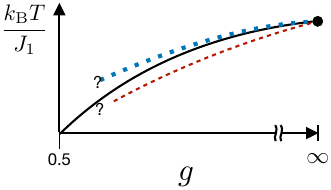}
\caption{(color online). Phase diagram at $g\geq 0.5$ according to the numerical results and theoretical arguments in the present work.  The entire phase boundary (black, solid) at $g>0.5$ is first-order.   Below the phase boundary is a line of pseudo critical points (red, dashed) associated with the spinodal line of the disordered phase, and above the phase boundary is a line of pseudo critical points (blue, dotted) associated with the spinodal line of the ordered phase.  The two pseudo critical lines asymptotically approach the phase boundary when $g\rightarrow\infty$; their fate when $g\rightarrow0.5^+$ remains unclear, though the data in Fig. \ref{fig:corrlengths}) suggests that the pseudo critical line associated with the disordered phase again approaches the phase boundary.
}
\label{fig:phasediagram2}
\end{figure}

However, this consensus overlooks the fact that the theoretical arguments in Ref. \cite{jin2012ashkin} do not rule out an alternative possibility that is also consistent with the numerical data: a line of weak first-order phase transitions at $0.67\lesssim g<\infty$ that \textit{asymptotically} turns into second-order transitions as $g\rightarrow\infty$. This is possible because a line of weak first-order phase transitions necessarily has a line of pseudo critical points on each side, and these three lines can asymptotically merge as $g\rightarrow\infty$; see Fig. \ref{fig:phasediagram2}.  The mapping onto the Ashkin-Teller model would then apply to one or both of the pseudo critical lines instead of the phase boundary itself.  Since weak first-order phase transitions can numerically appear as second-order phase transitions \cite{iino2019detecting}, the results \cite{jin2013phase,kalz2012location,murtazaev2015critical} indicating Ashkin-Teller criticality at $g\gtrsim0.67$ do not contradict this scenario.  When $g\rightarrow\infty$, the two pseudo critical lines in this scenario become a single critical point, whose central charge must be $c=1$ for consistency with the Ashkin-Teller criticality of one or both of the pseudo critical lines.  The central charge is a measure of the number of local degrees of freedom, and thus the two central charges of the pseudo critical lines do not add together at $g=\infty$ because the number of degrees of freedom does not change when $g\rightarrow\infty$.  This is consistent with the fact that at $g=\infty$ the model is two decoupled copies of the ordinary Ising model, each of which has central charge $c=1/2$, and each of which also has only half the number of degrees of freedom of the full model.

Below we present strong evidence that this alternative scenario is the actual case.  Further, we present strong evidence that the phase transitions in $0.5<g\lesssim0.67$ are also first-order.  These results comprise strong evidence for first-order phase transitions in this model; the evidence is strong instead of only suggestive because we take care to show that it is highly unlikely that the first-order signatures that we find are mere numerical artifacts.

Additionally, from the discussion above and from Fig. \ref{fig:phasetransitions} it seems plausible that there can be a first-order phase transition where the correlation length is very large in one phase but very small in the other; we find suggestive evidence that this is what occurs as $g\rightarrow0.5^+$ in this model.  Such a first-order phase transition can neither be categorized as a weak first-order phase transition nor as a strong first-order phase transition.  We are not aware of any previous suggestions in the literature of such a scenario.

In Sec. \ref{sec:numericalmethod} we explain the numerical method that we employ.  In Sec. \ref{sec:firstorder} we present numerical data that reveals the signatures of weak and anomalous first-order phase transitions.  In Sec. \ref{sec:imperfections} we present further data and theoretical arguments to show that the first-order signatures are not numerical artifacts.  In Sec. \ref{sec:discussion} we conclude with a discussion.

\section{Numerical Method}
\label{sec:numericalmethod}
The class of computational methods designed around tensor networks \cite{verstraete2008matrix,cirac2009renormalization,eisert2013entanglement,orus2014practical,bridgeman2017hand,orus2019tensor} is useful for studying both classical and quantum spin lattices; the following results from that field provide the basis for the numerical method of the present work:  Ref. [\onlinecite{orus2009first}] demonstrates with a quantum 2D spin lattice that adiabatic evolution of 2D tensor networks is able to reveal ground state energy density level crossings and order parameter discontinuities directly in the thermodynamic limit.  Ref. [\onlinecite{orus2008infinite}] demonstrates how to use a 1D tensor network, known as a matrix product state (MPS), to compute local observables directly in the thermodynamic limit of a classical 2D spin lattice.  This MPS approach amounts to a power method wherein the row-to-row transfer operator of the partition function is repeatedly applied to the MPS until the MPS is sufficiently converged toward the dominant eigenvector of the transfer operator; direct access to the thermodynamic limit is provided by presuming translation invariance up to a chosen unit cell size and computing with only a single unit cell.  The converged MPS may be interpreted as the groundstate wavefunction of a quantum 1D system that is dual to the classical 2D system \cite{suzuki1976relationship,zauner2015transfer}, and the correlations in the MPS may be interpreted as the bipartite entanglement in that wavefunction.  The entanglement capacity of an MPS is limited by the size of its matrices, and this approach therefore suffers from finite-entanglement effects instead of finite-size effects.

We use the tensor network representation of the partition function for the present model that is given in the Appendix of Ref. [\onlinecite{gangat2019phase}]. We compute the Helmholtz free energy density and order parameter with an MPS method similar to that in Ref. [\onlinecite{orus2008infinite}].

The converged MPS gives ready access to the Schmidt decomposition of the one-dimensional wavefunction that it represents: 
\begin{equation}
|\psi\rangle = \sum_{i=1}^{\chi} \lambda_i|\phi_i^L\rangle|\phi_i^R\rangle,
\label{eqn:schmidt}
\end{equation}
where $|\phi_i^L\rangle$ ($|\phi_i^R\rangle$) are states of the left (right) half of a bipartition of the system, $\lambda_i$ are real scalars in descending order, $\sum_i\lambda_i^2=1$, $\langle \phi_i^L|\phi_j^L\rangle=\langle \phi_i^R|\phi_j^R\rangle=\delta_{ij}$, and $\chi\in\mathbb{Z}^+$ (known as the \textit{bond dimension}) limits the entanglement capacity of the MPS and also determines the computational cost of its storage and manipulation.  A single application of the transfer operator to the MPS results in an increase of the bond dimension to $4\chi$, after which the Schmidt decomposition must be truncated (and renormalized) to keep the first $\chi$ terms only; the sum $\sum_{i=\chi+1}^{4\chi}\lambda_i^2$ is known as the \textit{truncation error}.

The converged MPS also gives ready access to the correlation length \cite{orus2014practical}.  While a more accurate method for computing the correlation length from the MPS exists \cite{rams2018precise}, the simple method in Ref. [\onlinecite{orus2014practical}] is sufficient for our goal here of arriving at only a qualitative understanding.

For each value of $g$, we perform independent adiabatic evolutions with fixed values of $\chi$ using a two-site unit cell (represented by an MPS with two tensors, one for each physical site).  For each adiabatic evolution in temperature at a given value of $g$, we first converge a randomly initialized MPS with the transfer operator at the starting temperature, compute the physical quantities with the converged MPS, then use that converged MPS as the initial MPS for convergence at the subsequent temperature; the rest of the adiabatic evolution is completed by using the converged MPS at each temperature as the initial MPS for the subsequent temperature.  The entanglement entropy, given by 
\begin{equation}
S=\sum_i \lambda_i^2 \textrm{log}(\lambda_i^2), 
\end{equation}
depends on both the local and nonlocal correlations in the MPS, and we therefore consider the MPS converged at a given temperature when  $S$  is converged to within a chosen tolerance ($\epsilon$).  Adiabatic evolutions in $g$ at fixed temperature are done in an analogous way.  This method becomes numerically exact in the limit of vanishing increments of $T$ or $g$, $\chi\rightarrow\infty$, and $\epsilon\rightarrow0$.  We note that a similar adiabatic evolution method for probing phase boundaries with MPSs is used in Refs. \cite{roberts2019deconfined,roberts2021one}, the difference being that their approach uses a variational method instead of a power method to converge the MPSs.

\section{Weak and anomalous first-order signatures}
\label{sec:firstorder}

The order parameter $(m_x,m_y)$ is defined by 

\begin{eqnarray}
m_x &= \lim_{N_\textrm{x} \to \infty}\frac{1}{N_\textrm{x}}\big|\sum_{x=1}^{N_\textrm{x}}(-1)^x\sigma_{(x,1)}\big|,\\
m_y &= \lim_{N_\textrm{y} \to \infty}\frac{1}{N_\textrm{y}}\big|\sum_{y=1}^{N_\textrm{y}}(-1)^y\sigma_{(1,y)}\big|, 
\end{eqnarray}
where $N_\textrm{x}$ ($N_\textrm{y}$) is the number of lattice sites in the $\hat{x}$ ($\hat{y}$) direction, and ($x,y$) are the coordinates of a spin.  The stripe-ordered phase can have either $~m_x>0$ and $~m_y=0$ or $~m_x=0$ and $~m_y>0$.

We check for Helmholtz free energy density crossings at $g=0.49$ but find none.  This by itself does not rule out the possibility of extremely weak first-order transitions at $g=0.49$; we return to this in Sec. \ref{sec:discussion}.

\begin{figure}[t]
\includegraphics[scale=0.43]{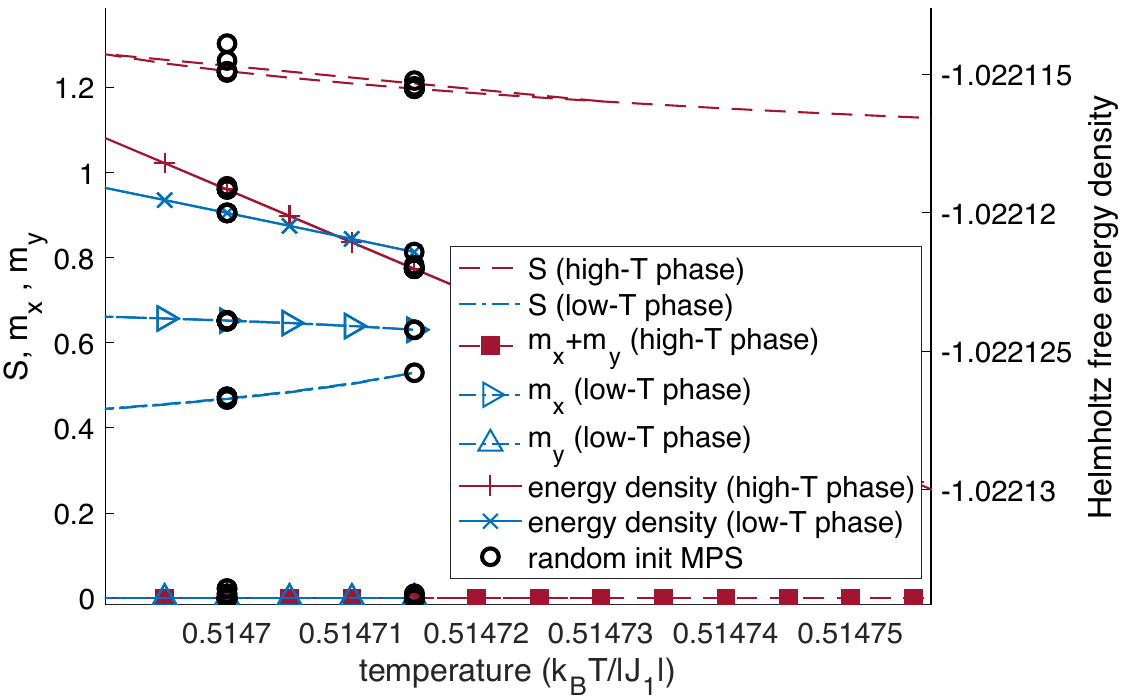}
\caption{(color online). Adiabatic evolutions in temperature at $g=0.51$ with MPS bond dimension $\chi=64$ and convergence tolerance $\epsilon=1$e-6.  $~m_y=0$ while $~m_x>0$ indicates that the low-to-high $T$ evolution remains in the stripe-ordered phase, and $m=0$ during the high-to-low $T$ evolution is consistent with the paramagnetic phase.  The Helmholtz free energy density crossing at a non-zero value of $~m_x$ signals a first-order phase transition.  After the endpoint of each adiabatic evolution is reached, the temperature is swept back over the crossing to confirm that contributions from imperfect adiabaticity are negligible.  The black circle data points are from convergence of ten randomly initialized MPSs at each of two temperatures, and confirm the presence of the first-order signatures that appear in the adiabatic data.  The monotonic increase of the entanglement entropy $S$ on the low-to-high $T$ and high-to-low $T$ paths suggests the proximity of the distinct pseudocritical points that are approached as the adiabatic evolution proceeds past the transition along each metastable branch; this is a generic feature of weak first-order phase transitions and is the mechanism behind their large correlation lengths \cite{fernandez1992weak,schulke2000dynamic} (see Fig. \ref{fig:phasetransitions}).}
\label{fig:g0.51_chi64_adiab_nonadiab}
\end{figure}

In Fig. \ref{fig:g0.51_chi64_adiab_nonadiab}) we illustrate the results of high-to-low $T$ and low-to-high $T$ adiabatic evolutions at $g=0.51$ with $\chi=64$ and $\epsilon=1$e-6.  The defining features of a first-order phase transition are apparent: a Helmholtz free energy density crossing occurs and the order parameter for the equilibrium state is discontinuous at the crossing.  

The apparent energy density crossing in Fig. \ref{fig:g0.51_chi64_adiab_nonadiab}) can not correspond to an energy density gap closing of a second-order transition for the following reasons: (1) As the gap-closing of a second-order transition is adiabatically approached from either side, the MPS must approach the same state and therefore the entanglement entropy ($S$) must approach the same value, whereas in Fig. \ref{fig:g0.51_chi64_adiab_nonadiab}) the two values of $S$ at the crossing differ greatly. (2) If the crossing is really a gap closing of a second-order transition, then the (quasi-)adiabatic evolution jumps from the ground state manifold to the excited state manifold as the crossing is traversed, which must result in a discontinuity in $S$ along either adiabatic path and a discontinuity in $(m_x,m_y)$ over the low-to-high $T$ path, but we find that both $S$ and $(m_x,m_y)$ are continuous over the energy density crossing along either adiabatic path. (3) A finite bond dimension imposes a finite correlation length \cite{orus2014practical} and therefore a finite gap, but there is no sign of a gap at the crossing point.

\begin{figure}[t]
\includegraphics[scale=0.62]{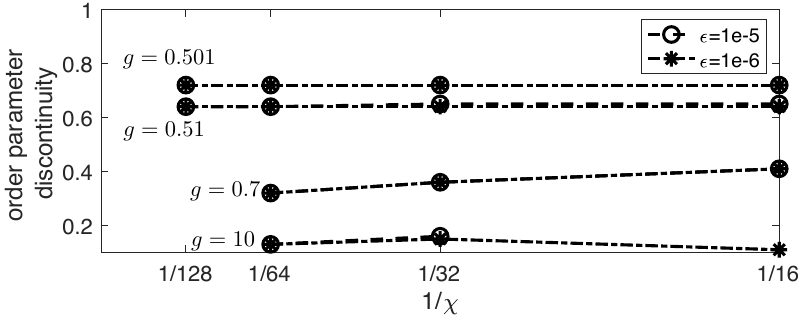}
\caption{The order parameter discontinuity (determined to 1e-2) at the Helmholtz free energy density crossing as a function of the inverse bond dimension.  Each data point is from independent adiabatic evolutions in $T$.  $\epsilon$ is the convergence tolerance for $S$ at each value of $T$.  The back and forth temperature sweeps across the energy crossing in some cases yield fluctuations in the order parameter discontinuity on the order of 1e-2, in which case the smallest value of the discontinuity is chosen.}
\label{fig:ordparamdiscontinuity}
\end{figure}

\begin{figure}
\subfloat[$g=0.55$]{{\includegraphics[width=0.47\textwidth]{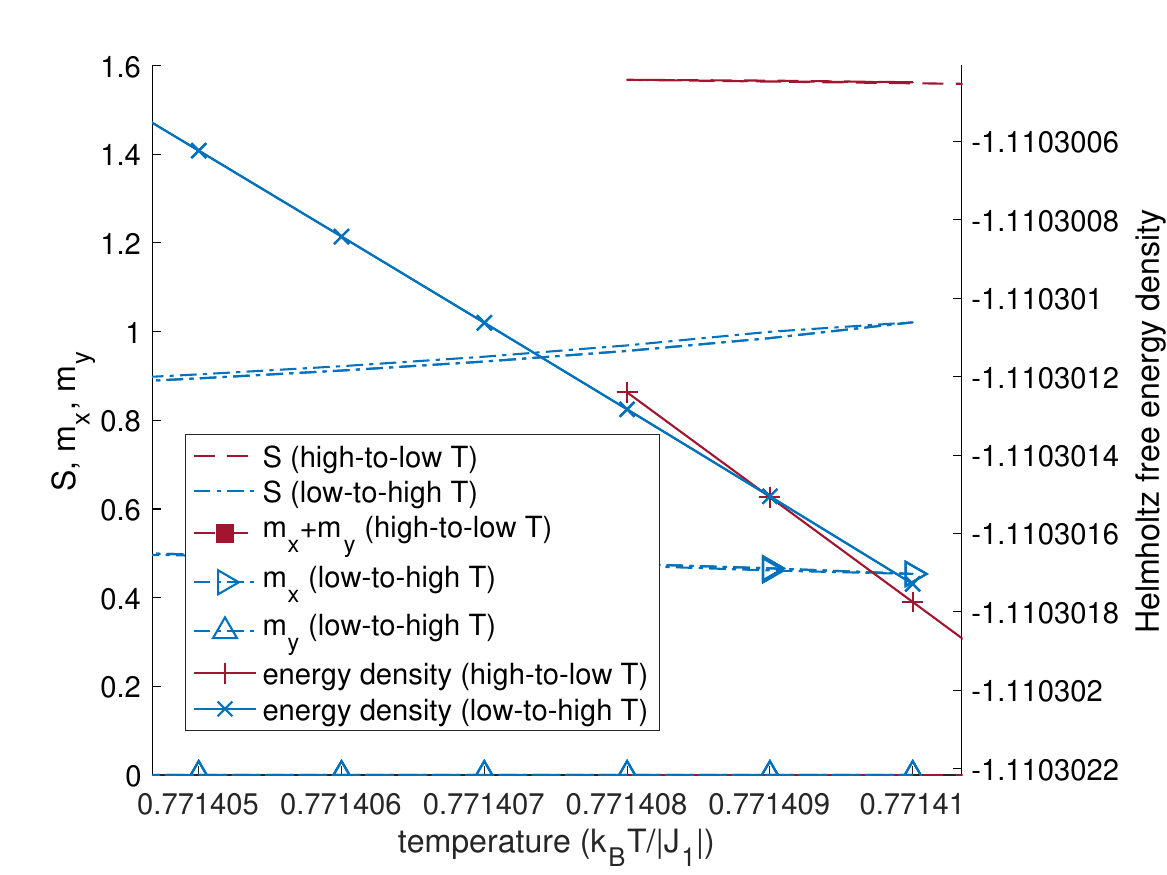} }} 

\subfloat[$g=0.6$]{{\includegraphics[width=0.47\textwidth]{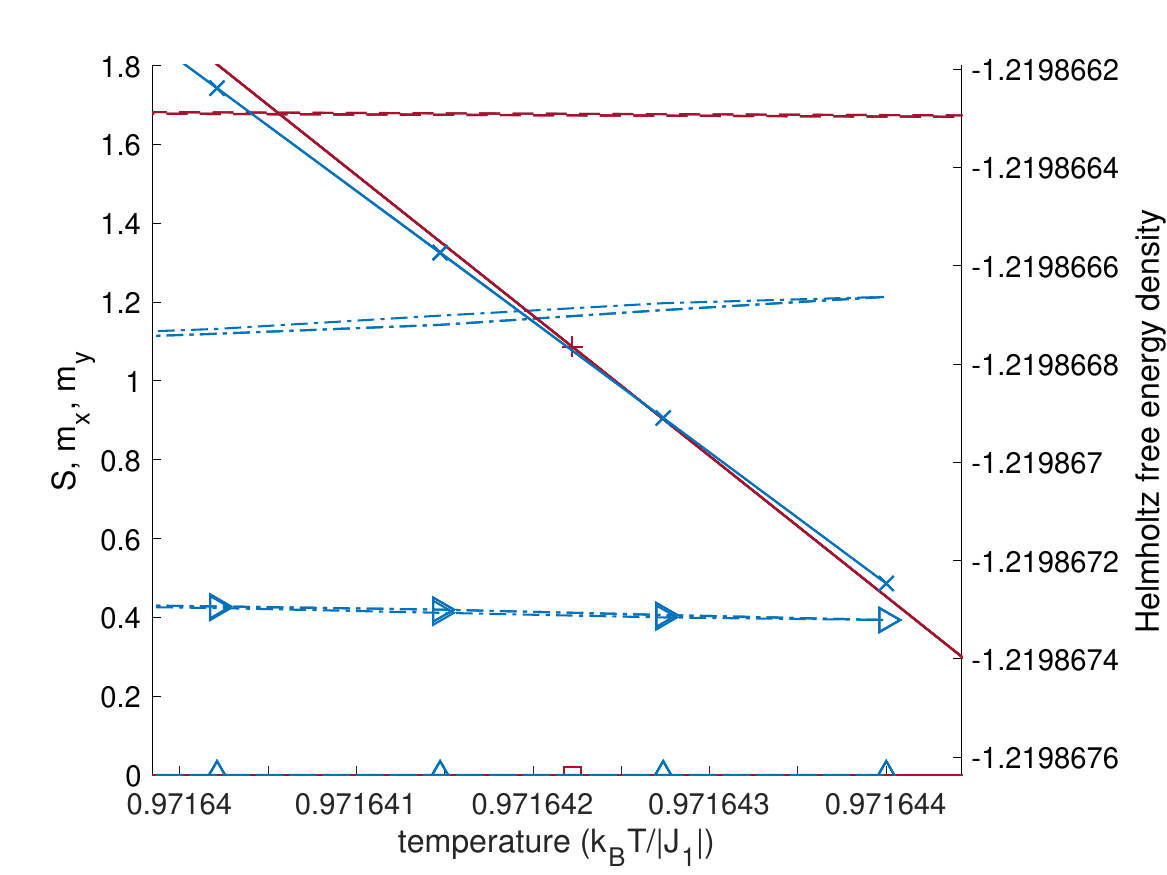} }}

\subfloat[$g=0.7$]{{\includegraphics[width=0.47\textwidth]{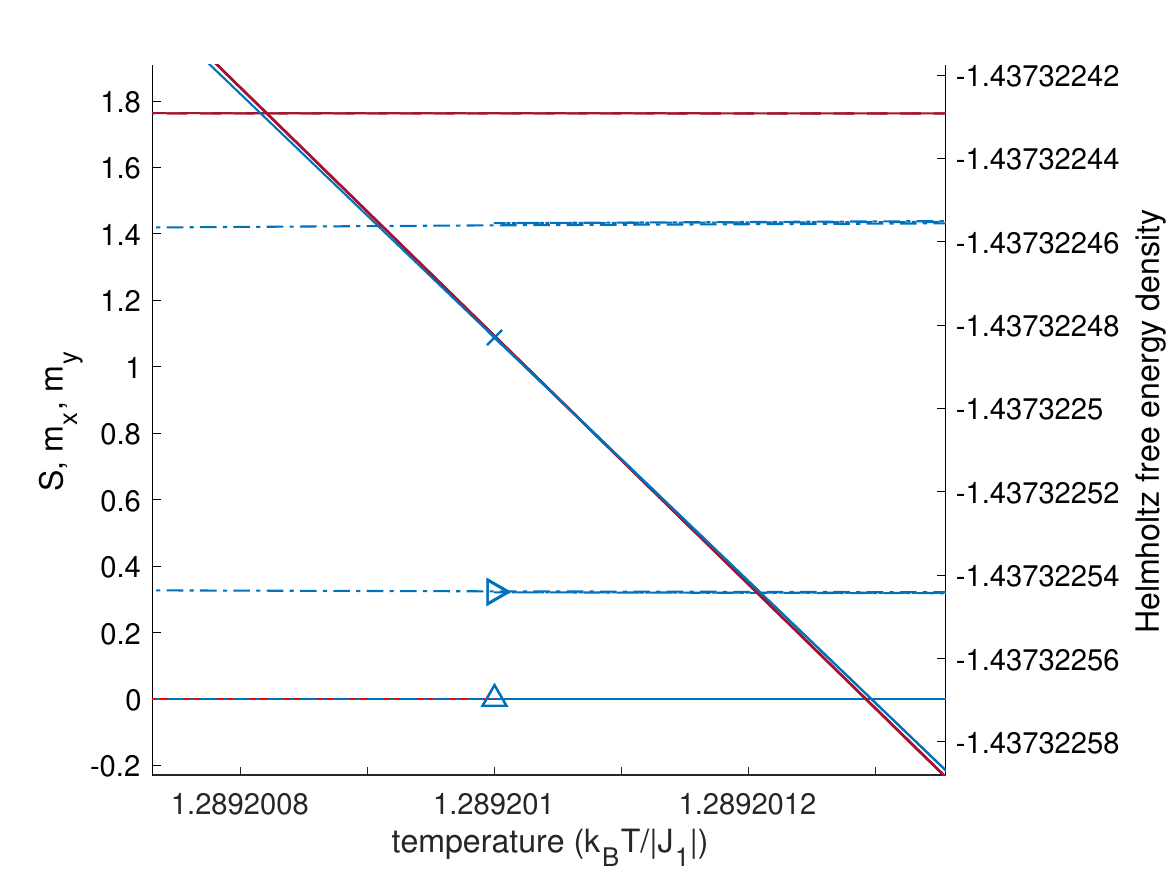} }}

\caption{(color online). Adiabatic evolutions in temperature at $g=0.55,~0.6,$ and $0.7$ with $\chi=64$ and $\epsilon\leq5$e-6 reveal signatures of first-order phase transitions. The adiabatic evolutions are swept back and forth over the free energy crossing to ensure that the crossing is not an artifact of imperfect adiabaticity. The order parameter discontinuity and the angle of incidence between the two Helmholtz free energy density curves decreases with increasing $g$. }
\label{fig:multig_chi64_adiab}
\end{figure}

We perform adiabatic evolutions with different values of $\chi$ and $\epsilon$ at a series of $g$.  With increasing $g$ at fixed $\chi$, the order parameter discontinuity (see Fig. \ref{fig:ordparamdiscontinuity}) and the angle of incidence between the two free energy curves (see Fig. \ref{fig:multig_chi64_adiab}) both decrease.  This is consistent with the requirement that the order parameter discontinuity and Helmholtz free energy crossing both disappear when $g\rightarrow\infty$.  Though the discontinuity at $g=0.7$ is not very small, this is still compatible with the possibility that previous numerics mistakenly find second-order transitions there; the weak first-order phase transition of the classical square lattice 5-state Potts model has a large order parameter discontinuity (about 0.49) \cite{igloi1999boundary}, yet appears as a second-order transition in certain finite-size numerics \cite{iino2019detecting} \footnote{A weak first-order transition may have a large order parameter discontinuity if the order parameter curve drops very steeply such that the energy crossing occurs at a large value of the order parameter yet is still very close to the pseudo critical point that lies at the zero point of the order parameter curve.}.

\begin{figure}
\includegraphics[scale=0.43]{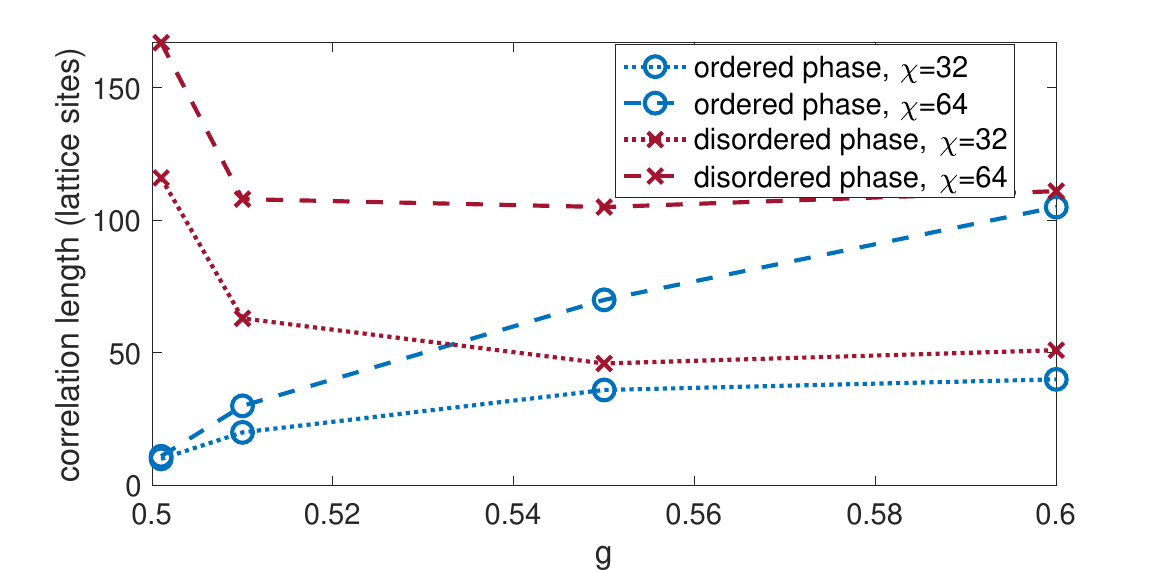}
\caption{(color online).  Correlation lengths at the first-order phase boundary at $0.501\leq g\leq0.6$ in the ordered and disordered phases as computed from adiabatic evolutions in $T$ with $\epsilon=5$e-6 and $\chi=32$ or $\chi=64$.  All of the correlation lengths computed here are much larger than the lattice spacing, which is consistent with a weak first-order scenario.  However, the trend in the data as $g\rightarrow0.5^+$ suggests an anomalous scenario wherein the correlation length is very large in only one of the phases at the first-order transition.}
\label{fig:corrlengths}
\end{figure}

\begin{figure}
\includegraphics[width=0.48\textwidth]{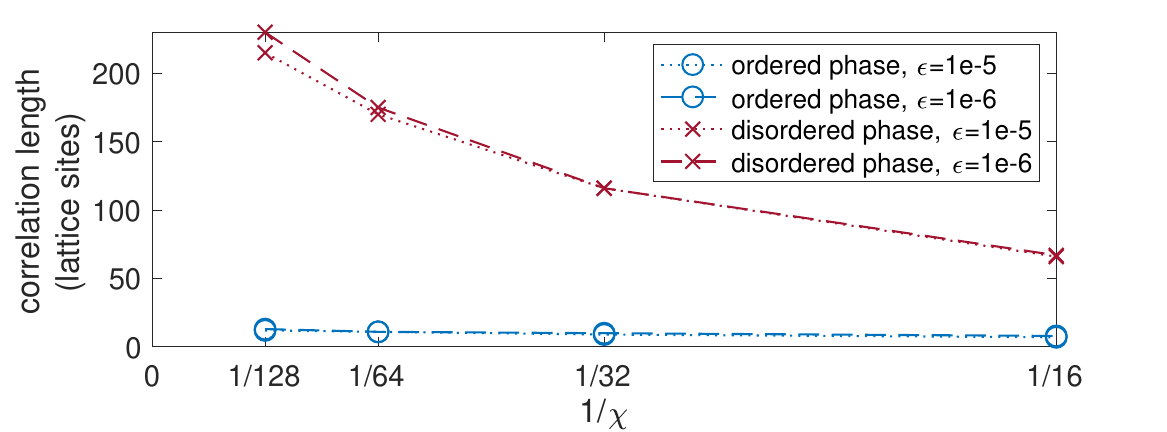}
\caption{(color online).  $g=0.501$. Correlation lengths at the first-order phase transition in the high-$T$ (disordered) and low-$T$ (ordered) phases.  The trend in the data as $\chi$ increases and $\epsilon$ decreases suggests the possibility at sufficiently small values of $g>0.5$ of an anomalous first-order phase transition wherein the correlation length is very large at the transition in one phase but very small in the other.}
\label{fig:g0.501_anomaly}
\end{figure}

\begin{figure*}
\subfloat[$g=1$]{{\includegraphics[width=0.47\textwidth]{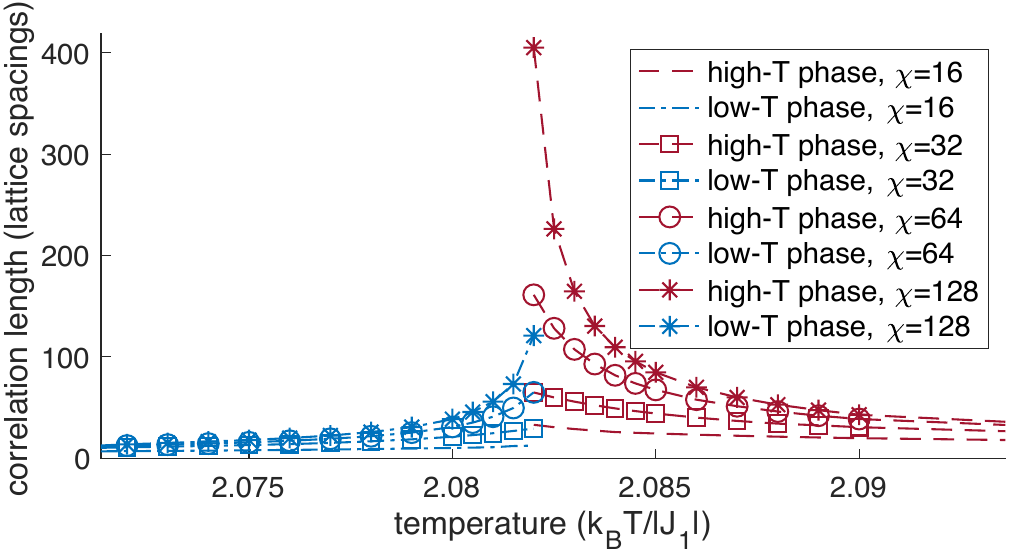} }} \hspace{0.4cm}
\subfloat[$g=10$]{{\includegraphics[width=0.47\textwidth]{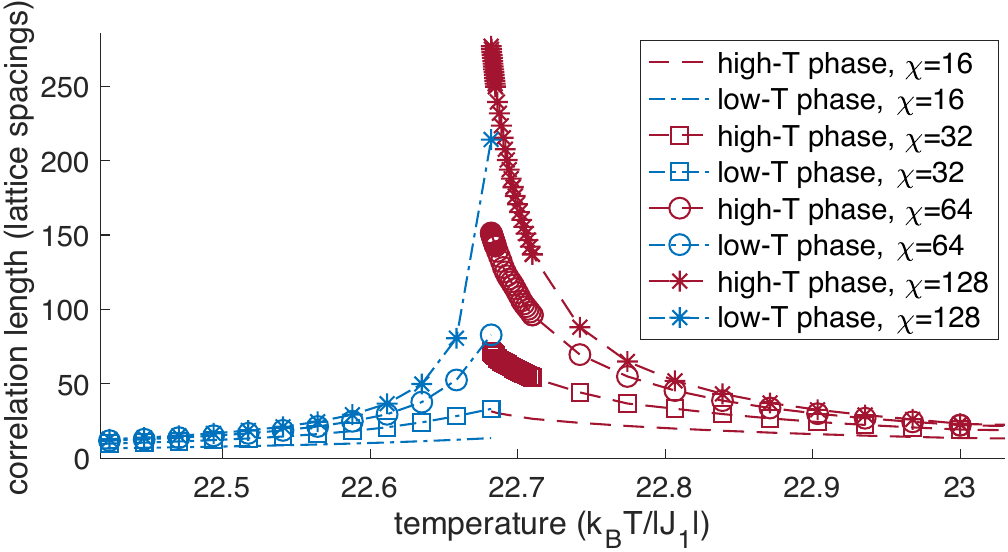} }}
\caption{(color online). Correlation lengths along adiabatic evolutions in $T$ toward the phase boundary with $\epsilon=1$e-4.  This value of $\epsilon$ is sufficient for good convergence of the correlation length data (see main text).  The trend in the data shows that the true correlation length in either phase at the transition is at least a few orders of magnitude larger than the lattice spacing, which is consistent with a weak first-order scenario that could easily be mistaken for a second-order scenario with other methods.  Sec. \ref{sec:imperfections} establishes that the apparent hysteresis is not a numerical artifact, and therefore we rule out a second-order transition at these values of $g$.}
\label{fig:corrlengths_g1_g10}
\end{figure*}

We also compute the correlation length in each phase near the phase boundary, as shown in Figs. (\ref{fig:corrlengths}), (\ref{fig:g0.501_anomaly}), and (\ref{fig:corrlengths_g1_g10}).   We obtain the data in Fig. \ref{fig:corrlengths_g1_g10}) with $\epsilon=1$e-4.  Converging $S$ to within this value of $\epsilon$ is sufficient for the correlation length data near the phase boundary because near criticality, $S\approx(c/6)\textrm{log}(\xi)$ \cite{calabrese2006entanglement}, where $c$ is the central charge and $\xi$ is the correlation length; $\epsilon=1$e-4 yields an error in $\xi$ of at most $\sim10$.

The large correlation lengths and their hysteresis that we find near the phase boundary are consistent with weak first-order transitions in the region $g\geq0.501$, and are also consistent with the \textit{numerical detection} of criticality in the region $0.67\lesssim g \lesssim1$ in Refs. \cite{kalz2012location,murtazaev2015critical} because, as mentioned above, weak first-order phase transitions can numerically appear as second-order phase transitions.  The hysteresis (assuming it is not a numerical artifact), however, is not consistent with \textit{actual} second-order transitions.  Interestingly, the data in Fig. \ref{fig:corrlengths}) suggests that as $g\rightarrow0.5^+$, a first-order transition arises that is neither strong nor weak but part of both: the correlation length is very large in one phase but very small in the other.  Further investigation of this is left for future work.

Thus, we find numerical evidence for first-order phase transitions in the region $0.5<g\leq10$.  This is at odds with the current consensus of second-order transitions in the region $g\gtrsim0.67$, but is consistent with the alternative possibility explained in Sec. \ref{sec:intro} of first-order transitions in the region $0.5<g<\infty$.  However, our numerical evidence necessarily contains contributions from numerical imperfections, such as finite bond dimension.  In the next section we show that such contributions are not responsible for the first-order signatures that we find, and thereby that our evidence for first-order phase transitions in the region $0.5<g\leq10$ is \textit{strong}.  We also explain why this leads to the conclusion that the first-order phase boundary only \textit{asymptotically} becomes second-order as $g\rightarrow\infty$.

\section{Consideration of numerical imperfections}
\label{sec:imperfections}

\subsection{Imperfect adiabaticity}
Here we address the possibility that the first-order signatures presented in Sec. \ref{sec:firstorder} are mere artifacts due to imperfect adiabaticity.  We note that on each adiabatic path we sweep the temperature back over the crossing after the extremal temperature is reached but we find only a negligible change.  We perform additional checks at $g=0.51$, $g=1$, and $g=10$:  At $g=0.51$ we converge ten randomly initialized MPSs at each of two temperatures, one above and one below the crossing, and find good agreement with the adiabatic data (see Fig.\ref{fig:g0.51_chi64_adiab_nonadiab}).  At $g=1$ and $g=10$ this check is computationally prohibitive due to the small angle of incidence between the two energy curves \footnote{We employ a power method, for which the rate of convergence to the equilibrium state is proportional to the energy gap.}, so we compare the results of adiabatic evolutions in $g$ (having fixed $T$) to the results of adiabatic evolutions in $T$ (having fixed $g$) such that both sets of evolutions meet at the same point in the phase diagram close to the phase boundary (either ($k_BT/|J_1|=2.082$, $g=1$) or ($k_BT/|J_1|=22.6821$, $g=10$)). In Appendix A we show that these complementary evolutions result in the same order parameter discontinuity and in the same values of $S$ at the point where they meet; this would be highly unlikely to occur if there were significant contributions from imperfect adiabaticity.   We therefore conclude that the first-order signatures that we find at $g>0.5$ are not an artifact of imperfect adiabaticity.

\subsection{Adiabatic evolutions with iTEBD}
Appendix B of Ref. \cite{tagliacozzo2008scaling} suggests that performing adiabatic evolutions with iTEBD itself can lead to the appearance of false metastabilities at second-order phase transitions, and that these metastabilities disappear if instead randomly initialized MPSs are used with iTEBD at each of the chosen points of the phase diagram.  Our data from randomly initialized MPSs rules out this possibility at $g=0.51$ (see Fig. \ref{fig:g0.51_chi64_adiab_nonadiab}).  Further, since such artifacts do not arise at the very strongly frustrated value of $g=0.51$, it is unlikely that they arise at larger values of $g$ (which have weaker frustration) when the same values of $\chi$ and $\epsilon$ are used.  Also, we note that the model at $g=10$ is only a perturbation of the model at $g=\infty$, which makes it unlikely that adiabatic evolutions with iTEBD produce artifactual energy crossing metastabilities at $g=10$ without doing so at $g=\infty$; our adiabatic evolutions at $g=0$ and $g=\infty$ (not shown) do not produce metastable energy crossings.  Finally, we note that though Appendix B of Ref. \cite{tagliacozzo2008scaling} presents data of hysteresis in the order parameter when adiabatically traversing the critical point of the quantum 1D Ising chain as evidence of artifactual metastability arising from adiabatic evolution with iTEBD, they do not show any metastable energy crossings, and their order parameter data is consistent with an alternative scenario: delayed spontaneous $Z_2$ symmetry breaking when evolving from the disordered phase into the ordered phase.  We therefore conclude that the first-order signatures that we find at $g>0.5$ are not due to the numerical imperfections that are specific to adiabatic evolutions with iTEBD.

\subsection{Finite bond dimension}
\label{sec:finitechi}
Here we address the possibility that the first-order transition signatures that we find at $g>0.5$ with finite $\chi$ are numerical artifacts that would disappear when $\chi\rightarrow\infty$.  

\subsubsection{Numerical evidence}
In Appendix B we show that near criticality, assuming conformal invariance, the entanglement entropy takes the form
\begin{equation}
S \approx -\frac{\nu c}{6}\textrm{log}|T-T_c| + constant,
\label{eqn:fit}
\end{equation}
and that this allows us to compute, at given values of $\chi$ and $\epsilon$, the value of $T_c$ via fitting.  The value of $T_c$ computed in this manner will differ between the two sides of the phase transition at finite $\chi$; for a second-order phase transition the two values will converge toward the same value with increasing $\chi$ and decreasing $\epsilon$, while for a first-order transition they will converge toward different values corresponding to the pseudo critical points.

\begin{figure}[b]
\includegraphics[scale=0.45]{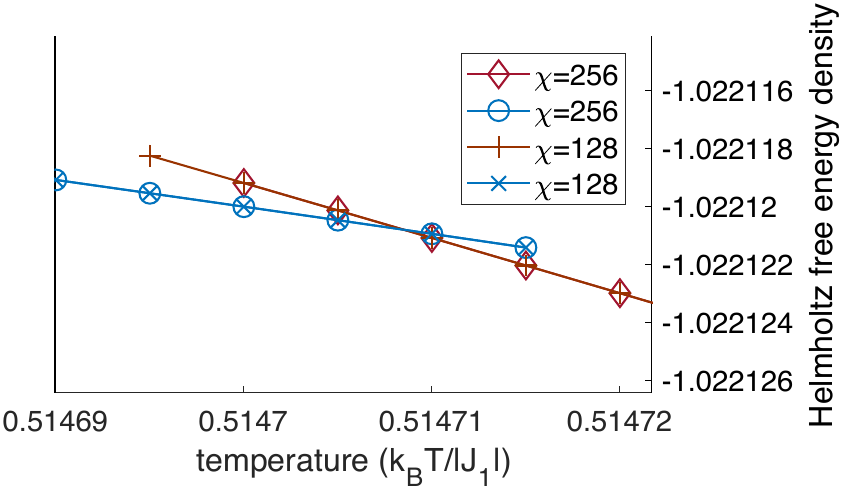}
\caption{(color online). $g=0.51$. Free energy density metastabilities ($\epsilon=1$e-6).  The metastable branches have the same position when independently computed with $\chi=128$ and $\chi=256$.  The entanglement entropy (not shown) is continuous along each adiabatic path, which rules out a second-order transition at these values of $\chi$.}
\label{fig:metastabilities_g0.51_chi128chi256}
\end{figure}

\begin{figure}[b]
\includegraphics[scale=0.45]{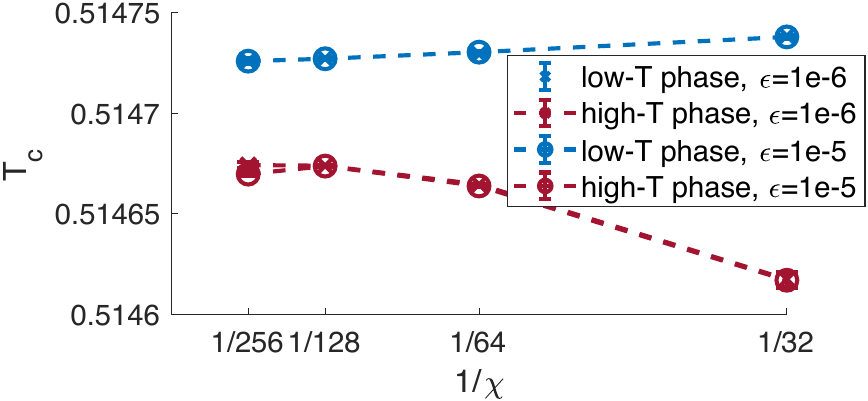}
\caption{(color online).  $g=0.51$. Least squares fitting of $T_c$ in Eq. (\ref{eqn:fit}) to the numerical data for $S$ as computed over adiabatic evolutions in temperature ($T$) on either side of the phase transition.  The error bars denote a $95\%$ confidence interval.  The values of $T_c$ from the high-$T$ and low-$T$ fittings converge to distinct values as $\chi$ is increased, which indicates that when $\chi\rightarrow\infty$ the transition consists of two pseudo critical points instead of a single true critical point.}
\label{fig:fittedTc_g0.51}
\end{figure}

\begin{figure}
\includegraphics[scale=0.43]{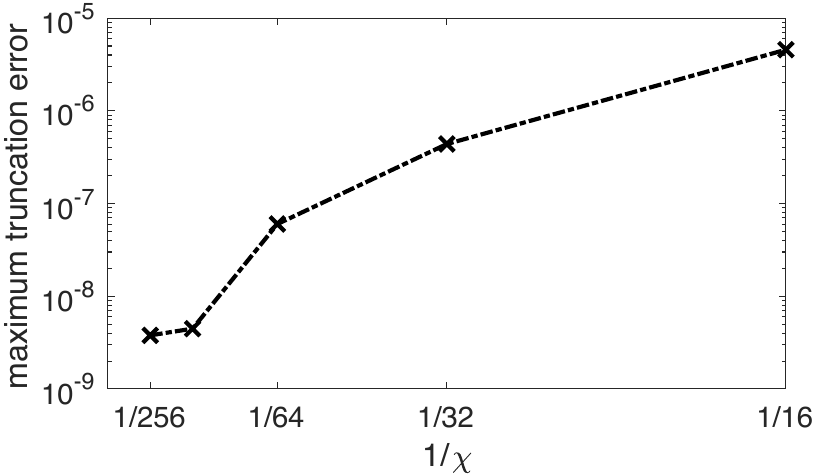}
\caption{$g=0.51$. Maximum truncation error of the converged MPS ($\epsilon=1$e-6) along the adiabatic evolutions across the phase transition (see Fig. \ref{fig:metastabilities_g0.51_chi128chi256}). The maximum is over the data from both the high-to-low-$T$ and low-to-high-$T$ adiabatic evolutions.  The evolutions are over the same temperature grids for all $\chi$.}
\label{fig:truncerror_g0.51}
\end{figure}

In Fig. \ref{fig:metastabilities_g0.51_chi128chi256} we show that the free energy density metastabilities that arise with adiabatic evolutions at $g=0.51$ do not change their location when $\chi$ is increased from $128$ to $256$.  Further, we show in Fig. \ref{fig:fittedTc_g0.51} that the fitted value of $T_c$ from the high-$T$ side and the fitted value of $T_c$ from the low-$T$ side converge to different values with increasing $\chi$, which is inconsistent with a second-order phase transition.  In Fig. \ref{fig:truncerror_g0.51} we show the corresponding truncation error (see Sec. \ref{sec:numericalmethod} for definition); the negligible truncation error at larger $\chi$ makes it unlikely that the data in Figs. \ref{fig:metastabilities_g0.51_chi128chi256} and \ref{fig:fittedTc_g0.51} would qualitatively change if $\chi$ was increased further.  The data in Fig. \ref{fig:fittedTc_g0.51} is also converged in $\epsilon$.  We take this as strong evidence that the phase boundary at $g=0.51$ remains first-order when $\chi\rightarrow\infty$.

\subsubsection{Theoretical arguments}
With increasing $g$, the pseudo critical points move closer together and it becomes untenable to assess the $\chi\rightarrow\infty$ limit through a purely numerical approach.  We therefore assess the $\chi\rightarrow\infty$ limit at larger $g$ by combining numerical data with the theory of finite entanglement scaling \cite{tagliacozzo2008scaling,pollmann2009theory,pirvu2012matrix,ueda2014doubling,corboz2018finite,czarnik2019finite} (also known as the theory of finite correlation length scaling), which provides a means of determining what happens in the $\chi\rightarrow\infty$ limit by using data computed with only moderate values of $\chi$.  For example, Ref. \cite{czarnik2019finite} demonstrates that the theory of finite entanglement scaling can be used with infinite projected entangled pair states (a type of two dimensional tensor network) to successfully determine properties of second-order phase transitions without resort to achieving convergence in the bond dimension.  This is analogous to how the theory of finite size scaling provides an understanding of the thermodynamic limit via computations with only modest system sizes.

We use the following results from the theory of finite entanglement scaling regarding how finite-$\chi$ effects in an MPS deform critical points: In a finite-$\chi$ MPS approximation of the ground state around a conformally invariant quantum 1D critical point, if $\chi$ is large enough to be within the finite-$\chi$ scaling limit (defined below), the divergence of the correlation length is replaced with a single finite peak \cite{tagliacozzo2008scaling}:
\begin{equation}
\xi_{\chi} \sim \chi^{\kappa},
\label{eqn:xi_chi}
\end{equation} 
where $\xi_{\chi}$ is the value of the said peak and $\kappa$ depends on the central charge ($c$) of the critical point \cite{pollmann2009theory,pirvu2012matrix}:
\begin{equation}
\kappa \approx \frac{6}{c\big(1+\sqrt{12/c}\big)}.
\end{equation} 
Further, the location ($\lambda^*_{\chi}$) of this peak evolves smoothly and monotonically as a function of $\chi$ such that it asymptotically goes to its true place ($\lambda^*$) on the phase diagram when $\chi\rightarrow\infty$ \cite{tagliacozzo2008scaling}:
\begin{equation}
\frac{|\lambda_{\chi}^* - \lambda^*|}{\lambda^*} \sim \chi^{-\kappa/\nu},
\label{eqn:shift}
\end{equation}
where $\nu$ is the critical exponent from the standard correlation length scaling relation that arises near criticality.  The finite-$\chi$ scaling limit is defined to be when $\xi_{\chi}\gg1$.  We note that the idea of a $\chi$-dependent correlation length originally appeared in the context of the corner transfer matrix renormalization group in Ref. \cite{nishino}.

As mentioned, when $g\rightarrow\infty$ the phase boundary of the present model necessarily becomes second-order.  This remains true with a finite-$\chi$ MPS representation: in the limit $g\rightarrow\infty$, the model becomes two decoupled copies of the classical 2D Ising model, the classical 2D Ising model is in the same universality class as the quantum 1D transverse field Ising model \cite{suzuki1976relationship,gang2006direct}, and the phase transition of the latter model does not develop energy crossings and order parameter discontinuities when represented by a finite-$\chi$ MPS that is computed with iTEBD \cite{tagliacozzo2008scaling}.  On the other hand, the data in Sec. \ref{sec:firstorder} shows that when $16\leq\chi\leq64$, the phase boundary in the region $0.501\leq g \leq10$ is first-order.  Therefore, when $16\leq\chi\leq64$ there is a value of $g$, denoted $g^*$, at which the phase boundary switches between first-order and second-order (see Fig. \ref{fig:phasediagram}).  Then the picture established for $16\leq\chi\leq64$ regarding the pseudo critical lines and the phase boundary is the one in Fig. \ref{fig:phasediagram2}, except that the location of $g^*$ is only known to be somewhere in $g>10$.

One way for the current consensus of $0.5\leq g^*\lesssim0.67$ to still remain true is for the entire phase boundary at $g>0.5$ to shift to lower values of $g$ as $\chi$ is increased from $64$ to $\infty$.  However, we can rule out this possibility because in Appendix C we show that, in accordance with Eq. (\ref{eqn:shift}), the phase boundary at $g\geq10$ monotonically shifts to larger values of $g$ when $\chi$ is increased above $64$.

Then the only way left for the current consensus of second-order transitions at $g\gtrsim0.67$ to remain true is if the two pseudo critical lines (see Fig. \ref{fig:phasediagram2}) fuse together as $\chi\rightarrow\infty$ such that $g^*$ moves to or below $0.67$.  However, we can rule out this possibility as well since it results in a contradiction: If the pseudo critical lines that appear when $\chi=64$ fuse together with increasing $\chi$ such that $g^*\lesssim0.67$ when $\chi\rightarrow\infty$, then there is a line of true critical points at $g\approx10$ when $\chi=\infty$ that must split into two pseudo critical lines when $\chi$ is reduced from $\infty$ down to $64$.  But in Appendix D we explain that since $\chi\geq64$ falls within the finite-$\chi$ scaling limit at $g=10$, Eq. (\ref{eqn:shift}) forbids critical points that occur at $g=10$ when $\chi=\infty$ to split into two pseudo critical lines as $\chi$ is reduced down to $64$.  We therefore have a lower bound of $g^*>10$ when $\chi\rightarrow\infty$.

Regarding the exact value of $g^*$, because the model at $g=10$ is far from the region of strong frustration and is only a perturbation of the model at $g=\infty$, it is unphysical that a switch of the phase boundary from first-order to second-order should occur at any finite value in $g>10$.  Therefore the only plausible scenario is that $g^*=\infty$ such that the first-order phase boundary only \textit{asymptotically} becomes second-order as $g\rightarrow\infty$, as depicted in Fig. \ref{fig:phasediagram2}.

\section{Discussion}
\label{sec:discussion}
First-order phase transitions have long been suspected to occur in this model, yet strong evidence has remained absent.  We addressed this in the following ways: We pointed out that the numerical data and theoretical arguments that form the basis for the current consensus of Ashkin-Teller criticality in the region $g\gtrsim0.67$ leave open the possibility of an alternative scenario: weak first-order phase transitions at $g\gtrsim0.67$ that asymptotically become second-order as $g\rightarrow\infty$.  We presented data and theoretical arguments to provide strong evidence that this alternative scenario is the true one, and further that the phase boundary is also first-order in the region $0.5<g\lesssim0.67$.  In contrast to previous studies that presented only 
\textit{suggestive} evidence for first-order transitions in this model, we established the first \textit{strong} evidence by taking care to account for contributions from numerical imperfections.  This was facilitated in part by the theory of finite entanglement scaling, which allowed us to make conclusions about the limit of infinite bond dimension ($\chi$) from data obtained with only modest $\chi$.  Additionally, we provided suggestive evidence that as $g\rightarrow0.5^+$, the first-order transitions become of a previously unobserved type that can be categorized as neither strong first-order nor weak first-order: the correlation length is very large in one coexisting phase and very small in the other.

The numerical correlation lengths at the phase boundary at $g>0.5$ are determined in part by the proximity of the pseudo critical lines to the phase boundary and in part by the value of $\chi$.  The locations of the pseudo critical lines shift with $\chi$, as required by Eq. (\ref{eqn:shift}).  Therefore, in order to confirm the anomalous first-order transitions at $g\rightarrow0.5^+$ with the present approach of adiabatic evolutions of MPSs, values of $g$ in the region $0.5<g<0.501$ must be investigated showing convergence with increasing $\chi$ of both the correlation length data and the location of the pseudo critical lines.  If the anomalous first-order transitions are confirmed in this model, it would be important to consider in which other models, both classical and quantum, they may also appear.

Regarding the pseudo critical lines themselves, while it is clear that they must merge with the phase boundary when $g\rightarrow\infty$, the following remains unclear: where the pseudo critical lines go when $g\rightarrow0.5^+$, which pseudo critical line exhibits the Ashkin-Teller criticality at $g\gtrsim0.67$, and what the criticality of either pseudo critical line is in the region $0.5<g\lesssim0.67$.  Recent results \cite{yoshiyama2023higher} indicate that at least one of the pseudo critical lines continues to have continuously varying critical exponents in the region $0.5<g\lesssim0.67$.  It may be possible to use the results in Refs. \cite{tagliacozzo2008scaling,pollmann2009theory,pirvu2012matrix} to compute the critical exponents and central charge along the pseudo critical lines with MPSs.

We left open the possibility of an extremely weak first-order transition at $g=0.49$. The fate of the pseudo critical lines at $g\rightarrow0.5^+$ has a bearing on this: If the pseudo critical line for the disordered phase (i.e., the pseudo critical line at lower $T$) merges with the phase boundary as $g\rightarrow0.5^+$, it would give an infinite correlation length to the phase transition at $g=0.5$ from the high-$T$ side, and would thereby support the idea of a second-order transition at $g=0.49$.

Finally, Ref. \cite{d2023diagnosing} showed good results for numerically detecting weak first-order phase transitions in other models with a different approach than the one used here. Future work may be able to use that approach as an independent check of the results that we have presented here.
\newline
\begin{acknowledgments}
We acknowledge Anders Sandvik for drawing this problem to our attention and for discussions and feedback.  We acknowledge Glen Evenbly for pointing out Ref. \cite{orus2009first}, for suggesting computation of pseudo critical point locations via fitting, and for technical help.  We acknowledge Wangwei Lan for technical help and discussions.  We also acknowledge discussions with Ying-Jer Kao, Ian McCulloch, Adam Iaizzi, André-Marie Tremblay, Thomas Baker, Ben Powell, Logan Wright, and Kai-Hsin Wu.  This research was partly supported by the Australian Research Council Centre of Excellence for Engineered Quantum Systems (EQUS, CE170100009).
\end{acknowledgments}

\bibliography{Ref}

\section*{Appendix A: hysteresis data demonstrating negligible contributions from imperfect adiabaticity}

\subsection{$g=1$}

\begin{figure*}
\subfloat[$g=1$]{{\includegraphics[width=0.47\textwidth]{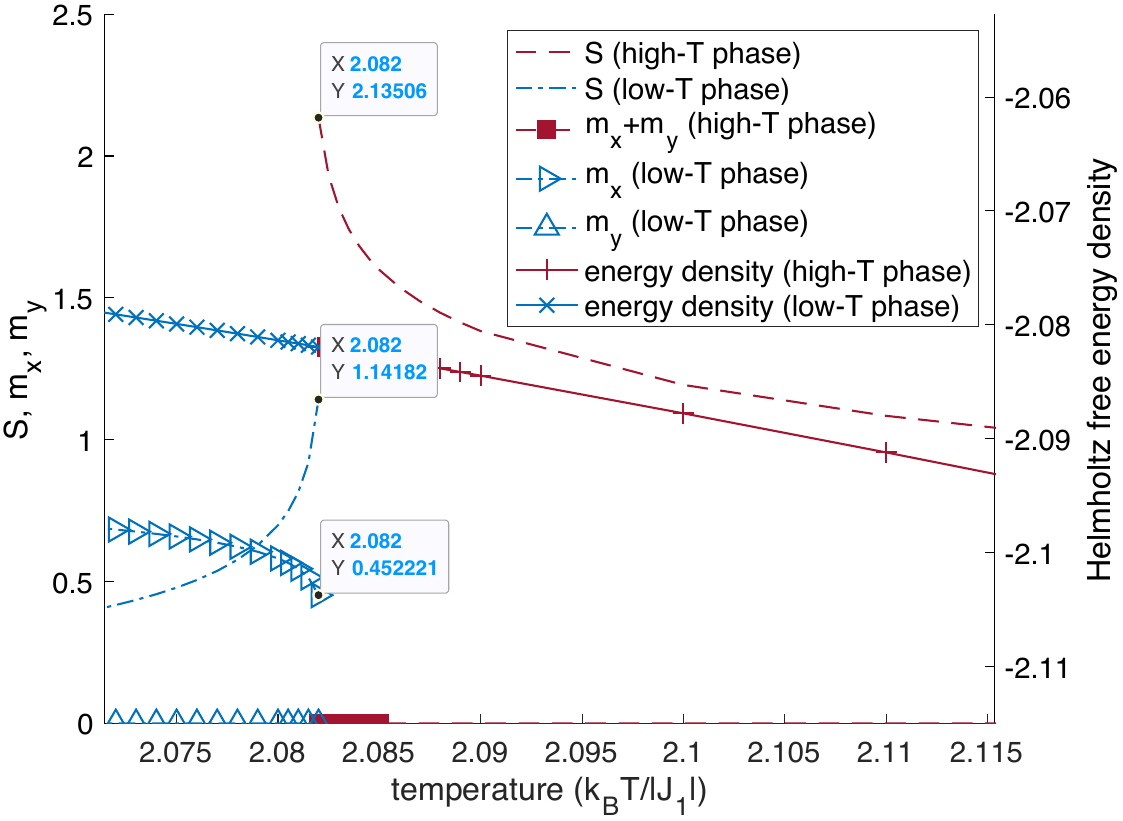} }} \hspace{0.4cm}
\subfloat[$k_BT/|J_1|=2.082$]{{\includegraphics[width=0.47\textwidth]{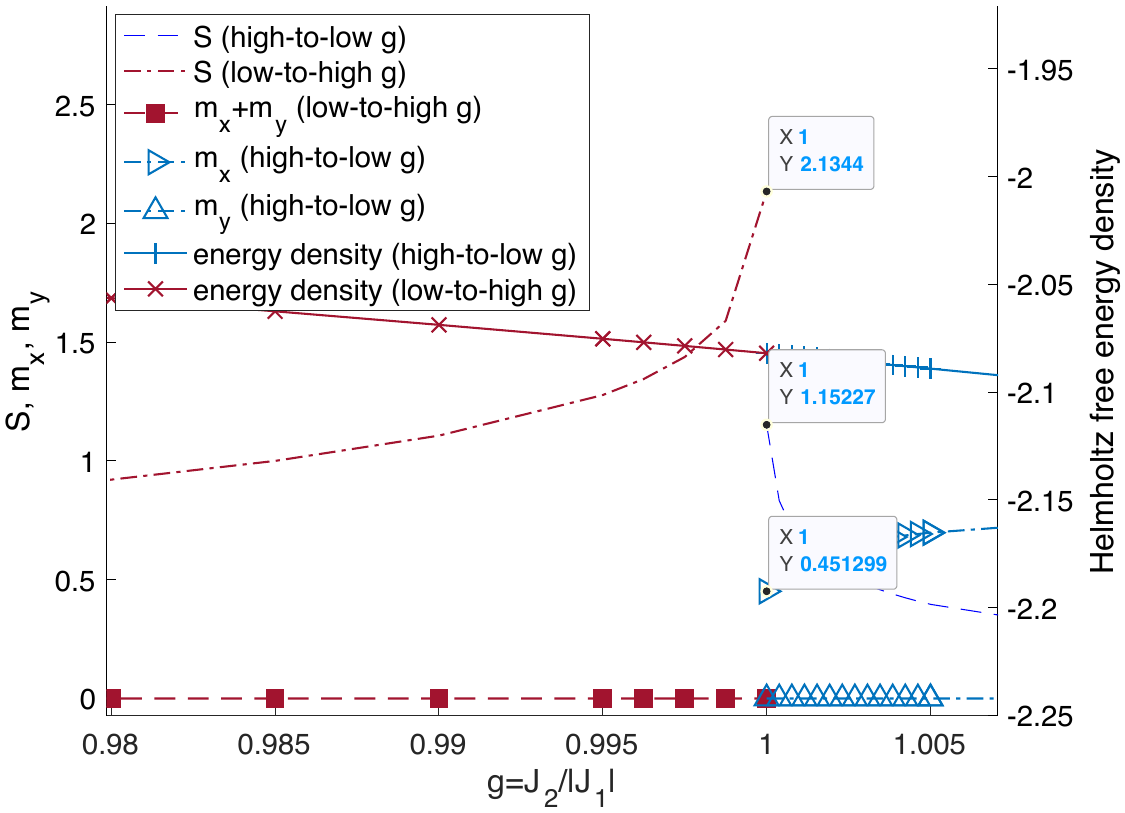} }}

\subfloat[$g=10$]{{\includegraphics[width=0.47\textwidth]{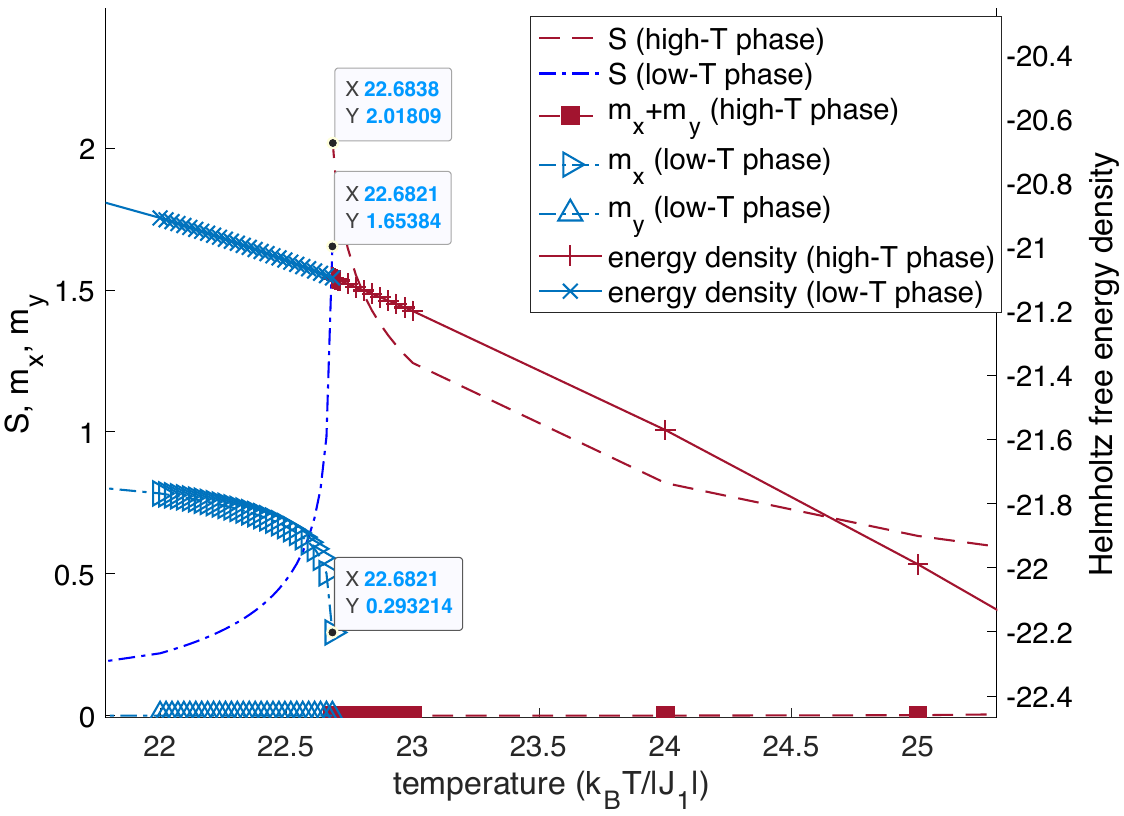} }} \hspace{0.4cm}
\subfloat[$k_BT/|J_1|=22.6821$]{{\includegraphics[width=0.47\textwidth]{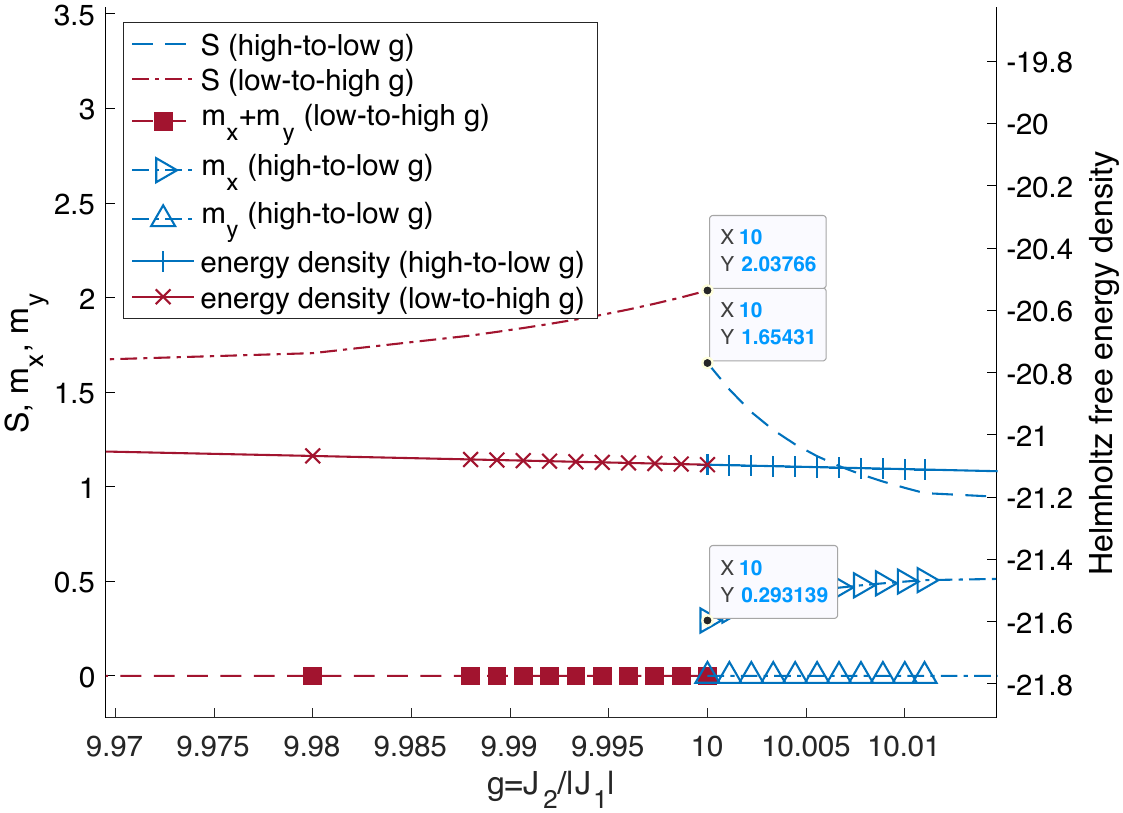} }}
\caption{(color online). $g=1$ (top) and $g=10$ (bottom): adiabatic evolutions ($\chi=128$, $\epsilon=1$e-4) in $T$ (left) and in $g$ (right) up to the vicinity of the phase boundary.  Hysteresis is apparent in both the entanglement entropy and the magnetization.  Agreement, within each phase, between the evolutions in $T$ and the evolutions in $g$ at their meeting points at $g=1$ and $g=10$ shows that imperfect adiabaticity does not contribute significantly to the hysteresis. The value of $\epsilon$ is much smaller than the difference in $S$ between the two phases at $g=1$ and $g=10$, indicating that the hysteresis in $S$ persists when $\epsilon\rightarrow0$.  Figs. \ref{fig:truncerr_g1} and \ref{fig:truncerr_g10} show that the truncation error is everywhere negligible. The correlation lengths along the evolutions in $T$ are shown in Fig. \ref{fig:corrlengths_g1_g10}.}
\label{fig:hysteresis_g1_g10}
\end{figure*}

\begin{figure}
\includegraphics[scale=0.43]{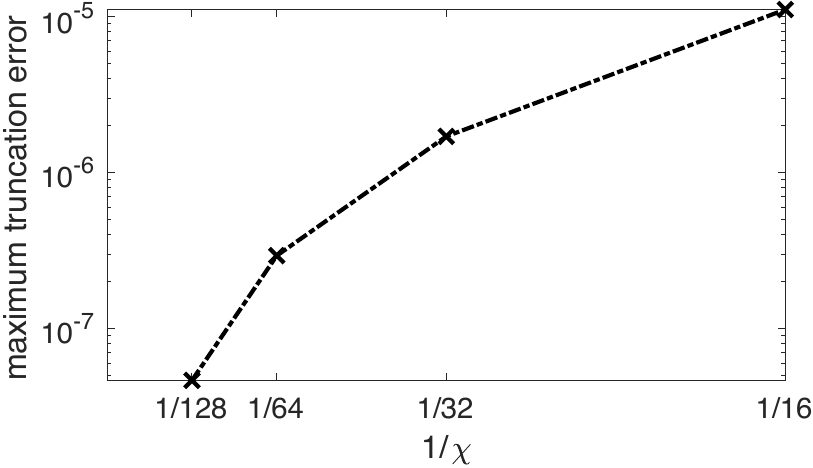}
\caption{$g=1$. Maximum truncation error of the converged MPS ($\epsilon=1$e-4) along the adiabatic evolutions in $T$ toward the phase boundary. The maximum is over the data from both the high-to-low $T$ and low-to-high $T$ adiabatic evolutions.  The evolutions are over the same temperature grids for all $\chi$.  The evolutions with $\chi=128$ are illustrated in the top panel of Fig. \ref{fig:hysteresis_g1_g10}.}
\label{fig:truncerr_g1}
\end{figure}

We use adiabatic evolutions in $T$ at $g=1$ and adiabatic evolutions in $g$ at $k_BT/|J_1|=2.082$ to observe hysteresis in $S$ (see Fig. \ref{fig:hysteresis_g1_g10}).  These evolutions all meet at the same point ($k_BT/|J_1|=2.082$, $g=1$) in the phase diagram.  The evolutions in the low-$T$ phase show agreement at that point, and the evolutions in the high-$T$ phase also show agreement at that point.  Since the evolutions in $T$ and the evolutions in $g$ are very distinct paths in the phase diagram, the errors that they accrue due to imperfect adiabaticity have very different origins.  Thus, the agreement at the point of intersection of the evolutions would be highly unlikely if the errors due to imperfect adiabaticity were appreciable.  We can therefore rule out the possibility that the hysteresis is merely an artifact of imperfect adiabaticity.  For completeness, in Fig. \ref{fig:truncerr_g1} we show truncation error data from these adiabatic evolutions.

\begin{figure}
\includegraphics[scale=0.43]{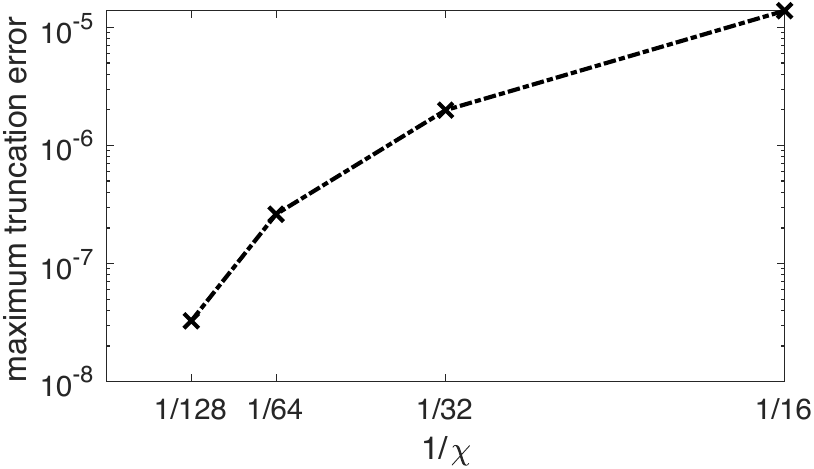}
\caption{$g=10$. Maximum truncation error of the converged MPS ($\epsilon=1$e-4) along adiabatic evolutions in $T$ toward the phase boundary. The maximum is over the data from both the high-to-low $T$ and low-to-high $T$ adiabatic evolutions.  The evolutions are over the same temperature grids for all $\chi$.  The evolutions with $\chi=128$ are illustrated in the bottom panel of Fig. \ref{fig:hysteresis_g1_g10}.}
\label{fig:truncerr_g10}
\end{figure}

\subsection{$g=10$}

The analysis here is analogous to the analysis above at $g=1$.  The common point on the phase diagram for the adiabatic evolutions here is $k_BT/|J_1|=22.6821$ and $g=10$.  The data in Fig. \ref{fig:hysteresis_g1_g10} shows that contributions from imperfect adiabaticity are negligible.  Truncation error data is presented in Fig. \ref{fig:truncerr_g10}.

\section*{Appendix B: pseudo critical point location via entanglement entropy fitting}
As mentioned in Sec. \ref{sec:numericalmethod}, the converged MPS at a given point in the phase diagram of the classical 2D system is an approximate representation of the ground state of the quantum 1D system that is dual to the classical 2D system at that point in the phase diagram.  For an infinite quantum 1D system near a conformally invariant critical point, Ref. \cite{calabrese2006entanglement} shows that the half chain entanglement entropy is given by
\begin{equation}
S = \frac{c}{6}~\textrm{log}~\xi_q, 
\end{equation}
where $c$ is the central charge and $\xi_q$ is the correlation length of the quantum 1D system.  Because of the duality, $c$ is the same between the quantum 1D system and its dual classical 2D system.  Near criticality we also have $\xi_q=\delta^{-\nu}$ and $\xi_{cl}=\tau^{-\nu}$, where $\xi_{cl}$ is the correlation length of the classical 2D system, $\delta$ and $\tau$ are small parameters, and the critical exponent $\nu$ is the same in both systems due to the duality.  More precisely, $\tau=|T-T_c|/T_c$, where $T_c$ is the temperature of the critical point, and the duality tells us that $\delta$ is a differentiable function of $\tau$.  Using a Taylor expansion around $\tau=0$ we can write $\delta(\tau)=\delta(0)+a\tau + \mathcal O(\tau^2)$, where $a$ is a constant.  Noting that $\xi_{q}=\xi_{cl}=\infty$ at the critical point (i.e., where $\tau=0$), we find $\delta(0)=0$.  Close to the critical point we can therefore write
\begin{equation}
S \approx -\frac{\nu c}{6}\textrm{log}|T-T_c| + constant.
\label{eqn:fit2}
\end{equation}

\begin{figure}[t]
\includegraphics[scale=0.43]{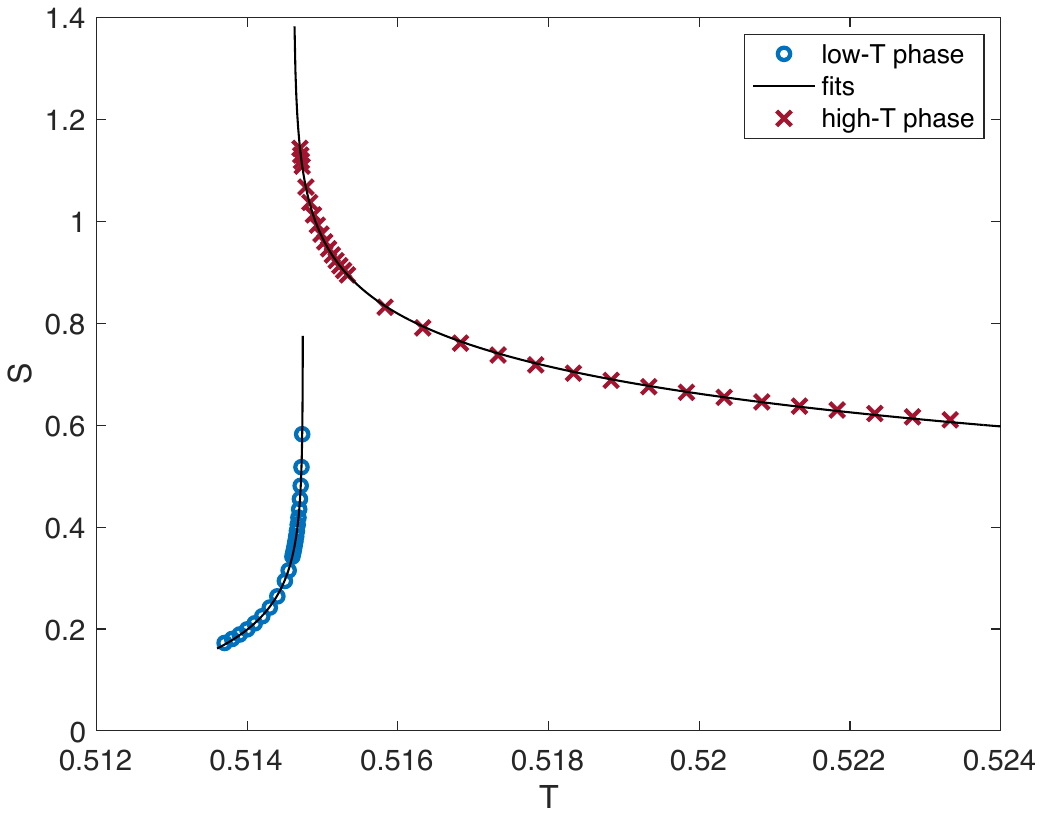}
\caption{(color online). $g=0.51$. Eq. (\ref{eqn:fit2}) fitting example ($k_B=J_1=1$). $S$ data is from adiabatic evolutions (both low-to-high $T$ and high-to-low $T$) toward the phase transition with $\chi=32$ and $\epsilon=1$e-6.  In this case, the fitted values of $T_c$ must be different since there is hysteresis in the data points.  Showing that they remain different when $\chi\rightarrow\infty$ (e.g., Fig. \ref{fig:fittedTc_g0.51}) rules out a second-order phase transition.}
\label{fig:fits}
\end{figure}

Since the half-chain entanglement entropy $S$ can be computed efficiently from the converged MPS, we may fit Eq. (\ref{eqn:fit2}) to the numerical data obtained by adiabatically approaching the phase boundary from both high and low temperature at fixed values of $g$.  Fig. \ref{fig:fits} shows an example.  For a continuous phase transition, both $\frac{\nu c}{6}$ and $T_c$ must be the same between the high- and low-temperature sides of the phase transition (though an exception can occur in anisotropic models, where it is possible that $\nu$ can be different between the two sides \cite{leonard2015critical}).  For a weak first-order phase transition, $T_c$ lies at the pseudo critical points and must therefore be different between the high- and low-temperature sides of the transition (see Fig. \ref{fig:phasetransitions}).  It is also possible for $\frac{\nu c}{6}$ to be different between the two sides of a weak first-order transition.

\begin{figure}[t]
\includegraphics[scale=0.43]{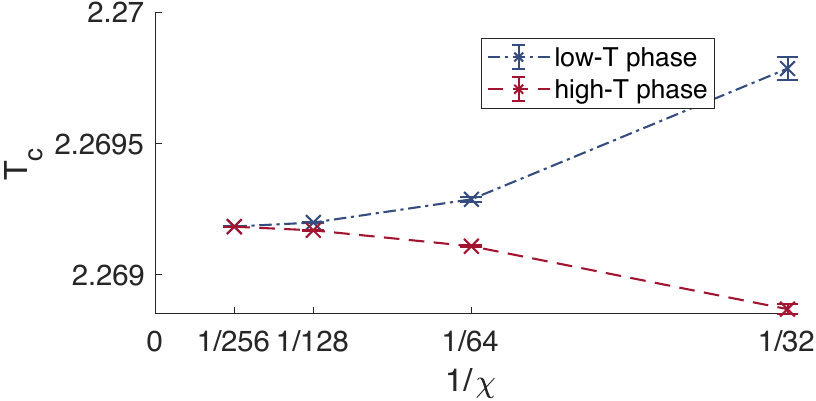}
\caption{(color online). Least squares fitting of $T_c$ in Eq. (\ref{eqn:fit2}) to the numerical data for the half-chain entanglement entropy ($S$) as computed over adiabatic evolutions in temperature ($T$) on either side of the known second-order phase transition at $g=0$ ($J_2=0$, $J_1=1$) with $\epsilon=1$e-6.  The error bars denote a $95\%$ confidence interval.  The values of $T_c$ from the high-$T$ and low-$T$ fittings converge toward each other as $\chi$ is increased, consistent with a second-order phase transition.  The numerical values are shown in Table \ref{tab:fittingdata}.}
\label{fig:fittedTc_g0}
\end{figure}

We use the known second-order phase transition at $g=0$ ($J_2=0$, $J_1=1$) as a benchmark case for the fitting of $T_c$ and $\frac{\nu c}{6}$.  We converge $S$ to within $1$e-$6$ at each temperature step.  We do not find good agreement between $\frac{\nu c}{6}$ (not shown) as extracted from fits of Eq. (\ref{eqn:fit2}) on the high- and low-temperature sides of the phase transition at $g=0$, but we do find agreement of $T_c$ to high precision (see Table \ref{tab:fittingdata} and Fig. \ref{fig:fittedTc_g0}).  We therefore only use fitted values of $T_c$ to draw conclusions at other values of $g$ (see Sec. \ref{sec:imperfections}).

\begin{table}[t]
\begin{tabular}{||c|c|c|c|c|c|c|c||}
\hline
$\chi$ & $T_c$ (low-$T$) & $T_c$ (high-$T$) & $\Delta T_c$ \\ \hline
32 & 2.269784411854  &  2.268866946007 & $\sim$9e-04\\ 
64  & 2.269285721313 &  2.269106864235 & $\sim$2e-04\\ 
128 & 2.269197257088  &  2.269167348571 & $\sim$3e-05\\ 
256 & 2.269181997444  & 2.269181205964 & $\sim$1e-06\\ \hline
\end{tabular}
\caption{$g=0$ fitting data ($k_B=J_1=1$). Least squares fitting of $T_c$ in Eq. (\ref{eqn:fit2}) to the numerical data for the half-chain entanglement entropy ($S$) as computed over adiabatic evolutions in temperature ($T$) at $J_2=0$, $J_1=1$ with $\epsilon=1$e-6. $\Delta T_c$ is the absolute difference between $T_c$ as fitted from the low-temperature side of the transition and $T_c$ fitted from the high-temperature side of the transition.  $\Delta T_c$ converges toward zero to high precision as $\chi$ is increased, which is consistent with the known second-order phase transition at this point in the phase diagram.  Fig. \ref{fig:fittedTc_g0} plots the fitted $T_c$ values with error bars.}
\label{tab:fittingdata}
\end{table}

\section*{Appendix C: monotonic shift of phase boundary at $g\geq10$ when $\chi\geq64$}
At $g=\infty$ the present system is two copies of the nearest neighbor classical Ising model on the square lattice, each of which has the Hamiltonian
\begin{equation}
H_{cl} = J_{cl}\sum_{\langle i,j \rangle} \sigma_i\sigma_j,
\label{eqn:classical2DHam}
\end{equation}
where $J_{cl}$ is a real scalar and the other variables are as defined in the Introduction, and whose critical temperature is given by \cite{kramers1941statistics}
\begin{equation}
T_c = \frac{2J_{cl}}{k_B \textrm{ln}(1+\sqrt{2})},
\label{eqn:Tc}
\end{equation}
where $k_B$ is Boltzmann's constant.

If we write the nearest neighbor quantum 1D Ising model with a transverse magnetic field as
\begin{equation}
H_q = - \sum_{\langle ij\rangle}\hat{\sigma}_i^z\hat{\sigma}_j^z -\lambda\sum_i\hat{\sigma}_i^x,
\label{eqn:quantum1DHam}
\end{equation}
then it has a quantum phase transition at $\lambda=\lambda^*=1$ \cite{pfeuty1970one}.

From Ref. \cite{sachdev2011} we know that there is a mapping between $H_{cl}$ and $H_{q}$ such that
\begin{equation}
\textrm{exp}(-2J_{cl})=\textrm{tanh}(\lambda J_{cl}).
\end{equation}
Thus, the value of $\lambda$ in Eq. (\ref{eqn:quantum1DHam}) fixes the value of $J_{cl}$ in Eq. (\ref{eqn:classical2DHam}). Further, the mapping is such that both Hamiltonians become critical simultaneously, which fixes the critical value of $J_{cl}$ at a value that we denote as $J_{cl}^*$:
\begin{equation}
\textrm{exp}(-2J_{cl}^*)=\textrm{tanh}(\lambda^* J_{cl}^*).
\end{equation}
As mentioned in Sec. \ref{sec:imperfections}, in a finite-$\chi$ MPS representation with sufficiently large $\chi$, the quantum 1D critical point is shifted to a value $\lambda^*_{\chi}$ so that:
\begin{equation}
\textrm{exp}(-2J_{cl,\chi}^*)=\textrm{tanh}(\lambda_{\chi}^* J_{cl,\chi}^*).
\label{eqn:mappingshift}
\end{equation}
Thus, combining Eqs. (\ref{eqn:Tc}) and (\ref{eqn:mappingshift}), we find that a monotonic shift in $\lambda_{\chi}^*$ as a function of $\chi$ results in a monotonic shift in $T_{c,\chi}$ as a function of $\chi$.

This establishes that the phase boundary at $g=\infty$ shifts smoothly and monotonically as a function of $\chi$ when $\chi$ is sufficiently large (i.e., within the finite-$\chi$ scaling limit).  The same must be true for the entire phase boundary at $g>0.5$, otherwise it would be unphysically crooked when $\chi\rightarrow\infty$.

\begin{figure*}
\subfloat[$\chi=32$, $k_BT/|J_1|=22.69535$]{{\includegraphics[width=0.47\textwidth]{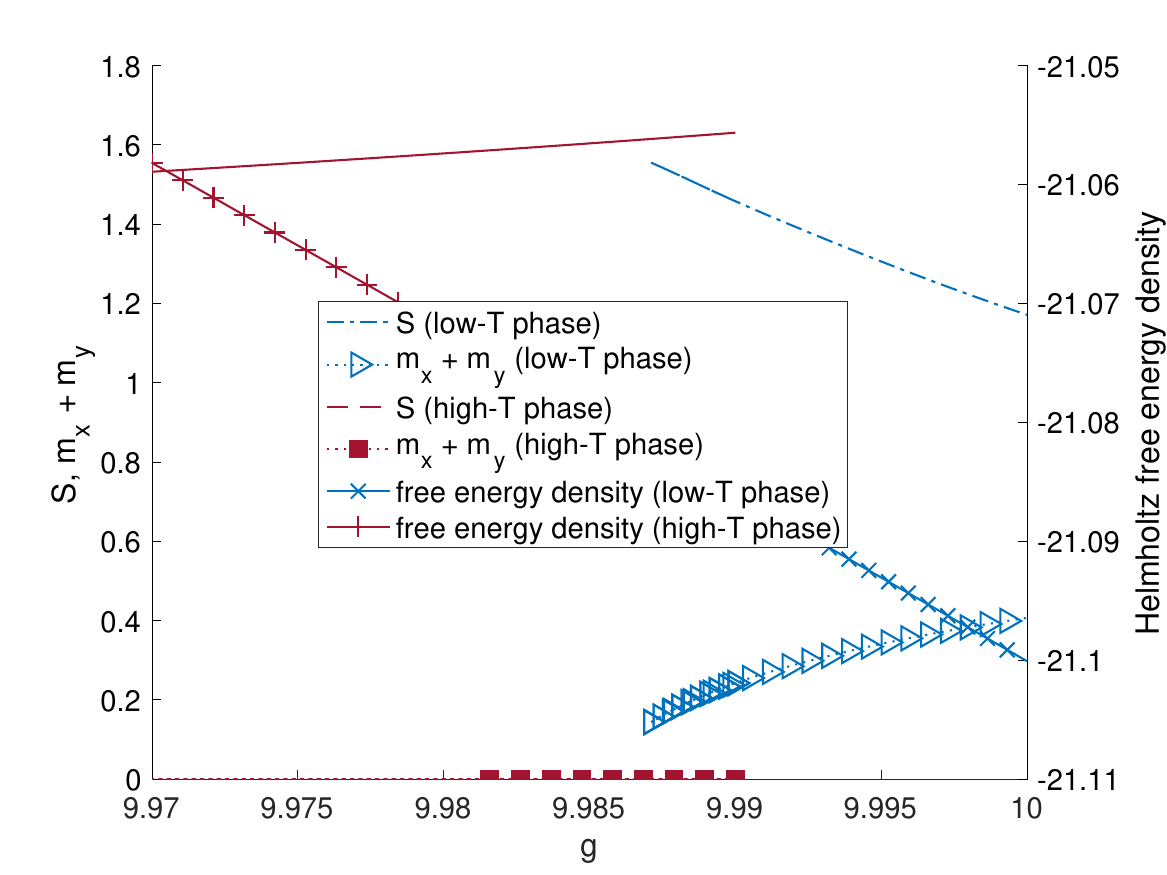} }} \hspace{0.4cm}
\subfloat[$\chi=32, g=10$]{{\includegraphics[width=0.47\textwidth]{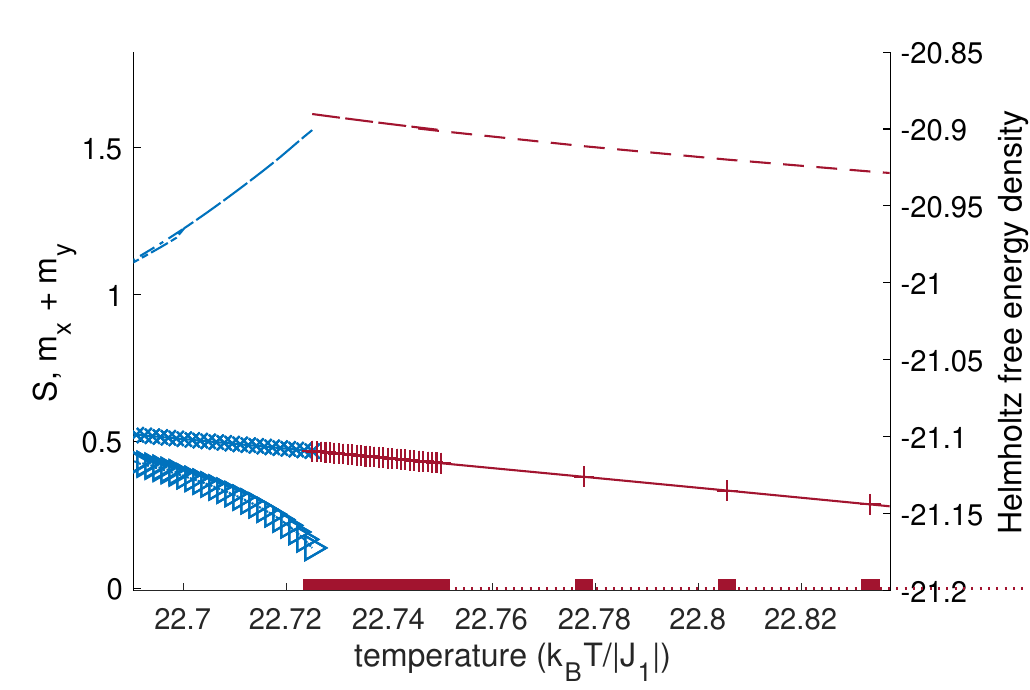} }}

\subfloat[$\chi=64$, $k_BT/|J_1|=22.69535$]{{\includegraphics[width=0.47\textwidth]{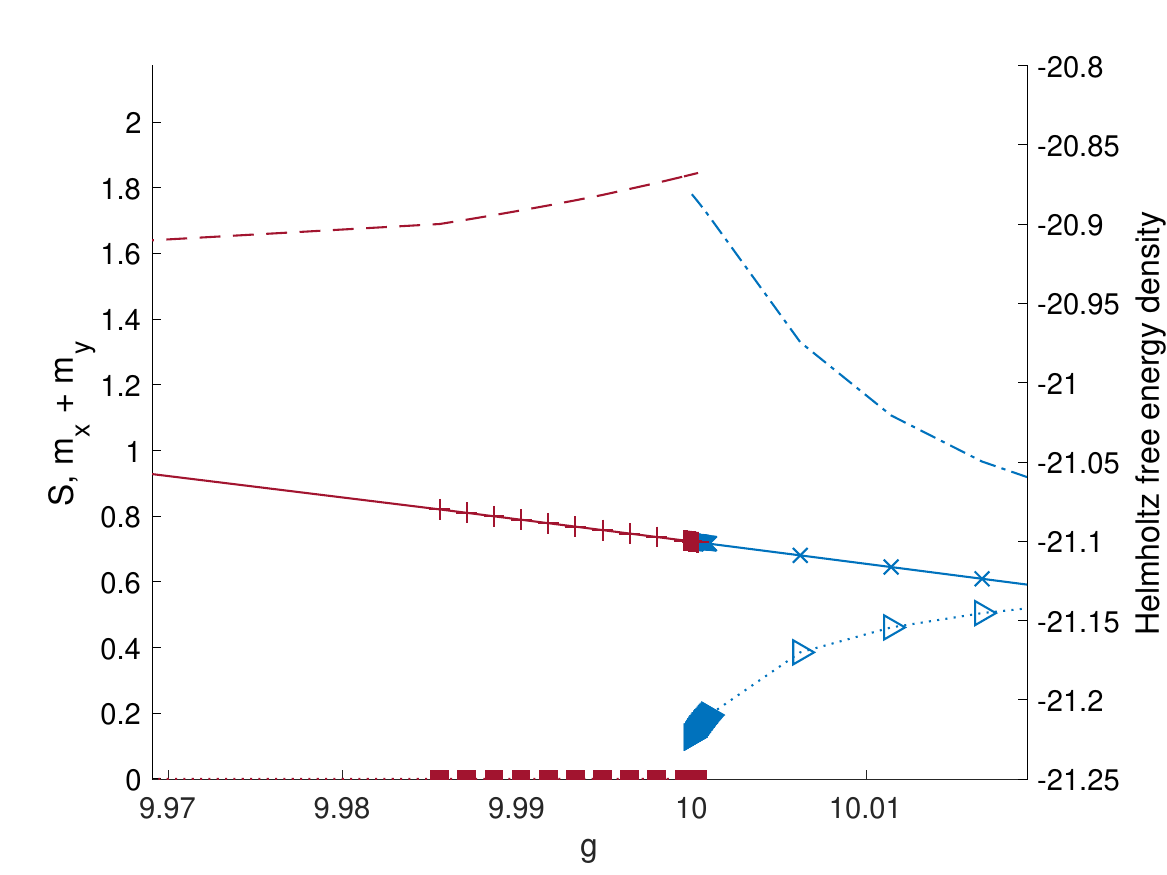} }} \hspace{0.4cm}
\subfloat[$\chi=64, g=10$]{{\includegraphics[width=0.47\textwidth]{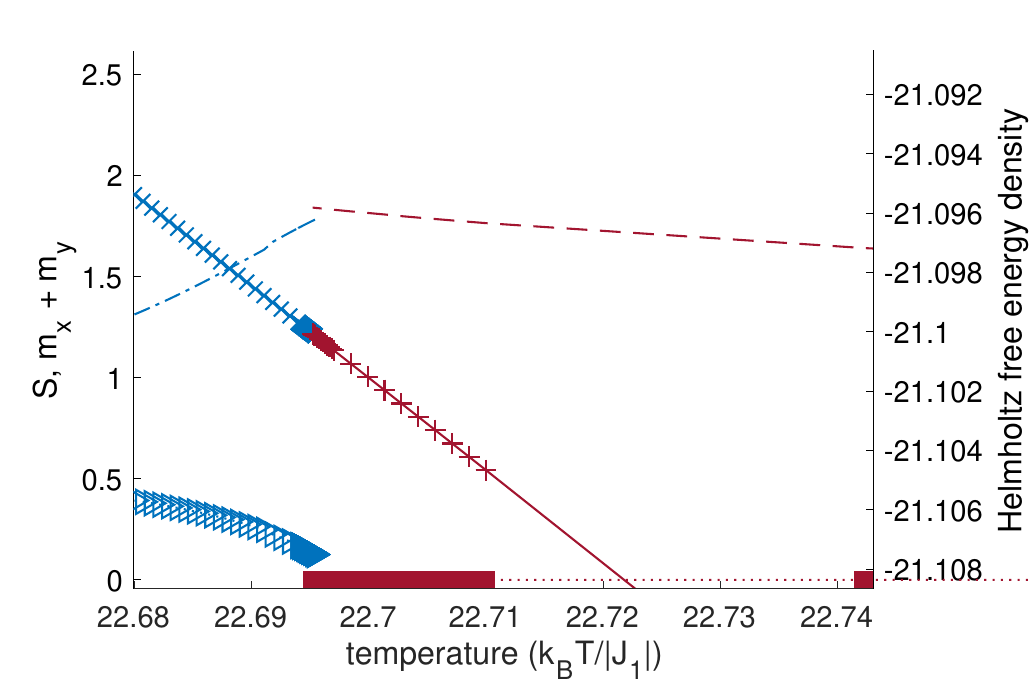} }}
\caption{(color online). Adiabatic evolutions (with $\epsilon=1$e-6) in $g$ at $k_BT/|J_1|=22.69535$ (left) and adiabatic evolutions in $T$ at $g=10$ (right) reveal that the phase boundary in the neighborhood of $g=10$ shifts to higher values of $g$ and lower values of $T$ with increasing $\chi$ (see also Fig. \ref{fig:crossingshift}).  The adiabatic evolutions are swept back and forth in the region of phase coexistence to ensure that contributions from imperfect adiabaticity are negligible.}
\label{fig:g10_chi32chi64_adiab}
\end{figure*}

\begin{figure*}
\centering
\subfloat[$k_BT/|J_1|=22.69535$]{{\includegraphics[width=0.48\textwidth]{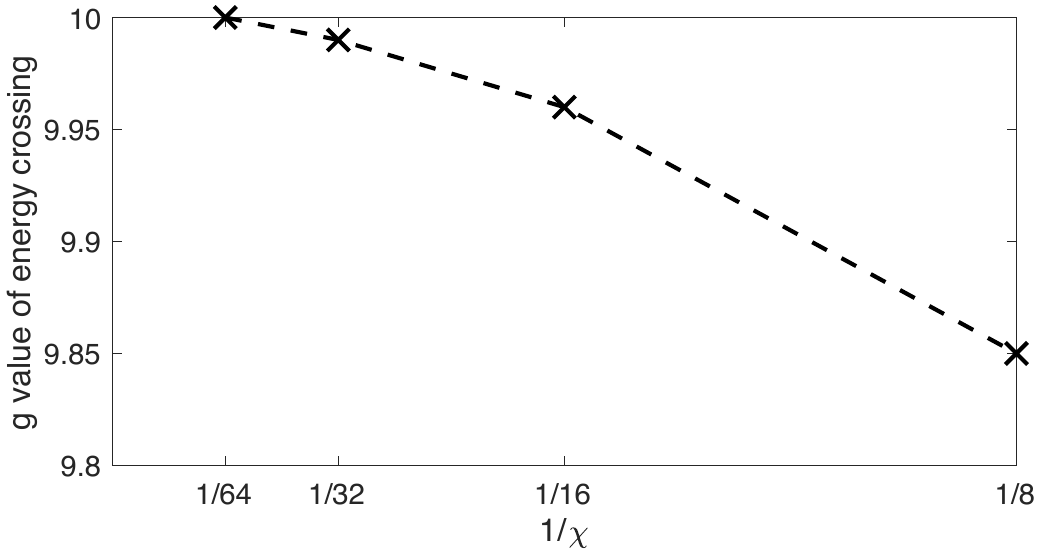} }} \hspace{0.4cm}
\subfloat[$g=10$]{{\includegraphics[width=0.48\textwidth]{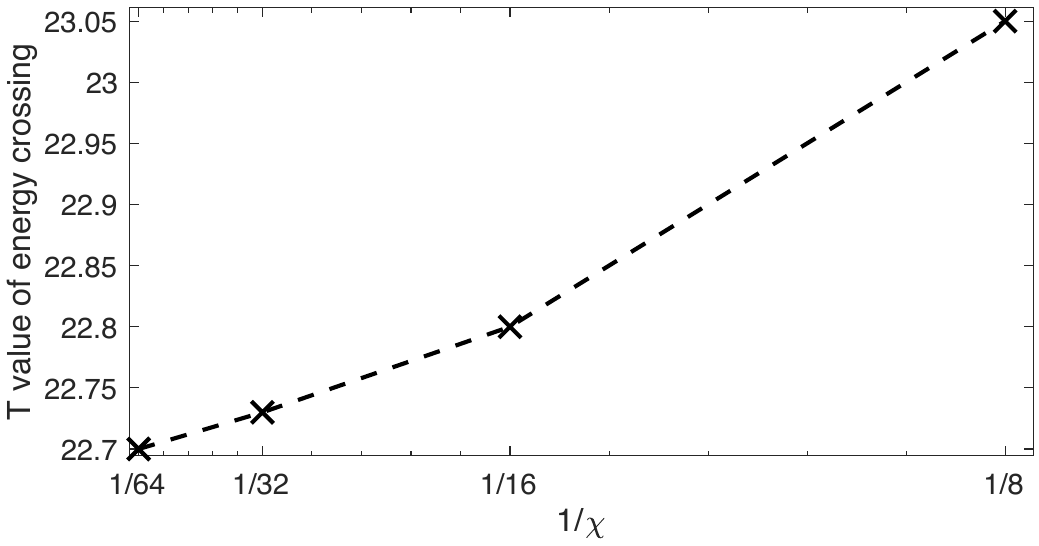}}}
\caption{Shift of the Helmholtz free energy density crossing near $g=10$ as a function of the inverse bond dimension as computed with adiabatic evolutions in either $T$ (right) or $g$ (left). All data here is from simulations with $\epsilon=1$e-6.  The phase boundary location (defined as the location of the energy crossing) shifts monotonically to lower values of $T$ and higher values of $g$ as $\chi$ is increased.}
\label{fig:crossingshift}
\end{figure*}

The correlation length data in Fig. \ref{fig:corrlengths_g1_g10} shows that $\chi\geq64$ falls within the finite-$\chi$ scaling limit at $g\geq10$.  This is consistent with the fact that the model at $g=10$ is only a perturbation of the model at $g=\infty$, and at $g=\infty$ the monotonic shift of the critical point with $\chi$ that is required by Eq. (\ref{eqn:shift}) is valid all the way down to $\chi=2$ \cite{tagliacozzo2008scaling}.  It is also consistent with the numerical data in Figs. \ref{fig:g10_chi32chi64_adiab} and \ref{fig:crossingshift}, which shows that the phase boundary around $g=10$ monotonically shifts, in agreement with Eq. (\ref{eqn:shift}), to larger values of $g$ and smaller values of $T$ as $\chi$ is increased from $16$.  Also, it is consistent with Eq. (\ref{eqn:xi_chi}) in the following sense:  At $g=\infty$ we have exactly $\xi_\chi = \chi^{\kappa}$ \cite{tagliacozzo2008scaling}, so we expect the prefactor $a$ in $\xi_\chi = a\chi^{\kappa}$ at $g=10$ to be close to unity since the model at $g=10$ is only a perturbation of the model at $g=\infty$.  If $g=10$ has a second-order phase transition, then previous results \cite{jin2012ashkin,jin2013phase,kalz2012location,murtazaev2015critical} establish that it has central charge $c=1$, which yields $\kappa\approx1.344$.  Then at $g=10$ we easily satisfy $\xi_\chi \gg 1$ when $\chi\geq64$.

We therefore conclude that the entire phase boundary at $g\geq10$ monotonically shifts to larger values of $g$ as $\chi$ is increased above $64$.

\section*{Appendix D: non-splitting of critical lines lines at $g\geq10$ when $\chi\geq64$}
Eq. (\ref{eqn:shift}) estalishes that conformally invariant quantum 1D critical points that appear within the finite-$\chi$ scaling limit (i.e., when $\chi$ is large enough so that $\xi_{\chi}\gg1$) can not split into two pseudo critical points as $\chi$ is varied within the finite-$\chi$ scaling limit.  The results in Ref. \cite{suzuki1976relationship} establish that the critical points of a given classical 2D spin lattice are equivalent to the critical points of some quantum 1D spin lattice.  Therefore, conformally invariant critical points that appear in the present classical 2D model within the finite-$\chi$ scaling limit also can not split into two pseudo critical points as $\chi$ is varied within the finite-$\chi$ scaling limit.

If the present classical 2D model has a line of critical points at $g\geq10$ when $\chi=\infty$, then Refs. \cite{jin2012ashkin,jin2013phase,kalz2012location,murtazaev2015critical} establish that it must have conformal invariance with $c=1$.  Also, the correlation length data in Fig. \ref{fig:corrlengths_g1_g10} reveals $\xi_{\chi}\gg1$ at $g=10$ when $\chi\geq64$, so $\chi\geq64$ lies within the scaling limit at $g=10$.  Therefore, Eq. (\ref{eqn:shift}) must apply to any line of critical points in the region $g\geq10$ when $\chi\geq64$.  This is consistent with the fact that $g=10$ is only a perturbation of the model at $g=\infty$, the critical point of the model at $g=\infty$ maps, via the classical-quantum correspondence \cite{suzuki1976relationship}, on to the critical point of the quantum 1D Ising model, and Eq. (\ref{eqn:shift}) is valid for the latter all the way down to $\chi=2$ \cite{tagliacozzo2008scaling}.  But Eq. (\ref{eqn:shift}) forbids a critical point from splitting into multiple pseudo critical points.  It is therefore forbidden for a line of critical points that appear at $g\geq10$ when $\chi=\infty$ to split into two pseudo critical lines as $\chi$ is reduced down to $64$.

\end{document}